\documentclass[twocolumn,aps,superscriptaddress,showpacs]{revtex4}
\usepackage{hyperref}
\usepackage{amssymb}
\usepackage{amsmath}
\usepackage{graphicx}
\usepackage[normalem]{ulem}
\usepackage{url}
\usepackage{color}

\begin{document}
\title{
Comparison of Heavy-Ion Transport Simulations: Mean-field Dynamics
in a Box\\
}

\author{Maria Colonna}
\email{colonna@lns.infn.it}
\affiliation{INFN-LNS, Laboratori Nazionali del Sud, 95123 Catania,
Italy}

\author{Ying-Xun Zhang}
\email{zhyx@ciae.ac.cn}
\affiliation{China Institute of Atomic Energy, Beijing 102413, China}
\affiliation{Guangxi Key Laboratory Breeding Base of Nuclear Physics and Technology, Guilin 541004, China}

\author{Yong-Jia Wang}
\email{wangyongjia@zjhu.edu.cn}
\affiliation{School of Science, Huzhou University, Huzhou 313000,
China}

\author{Dan Cozma}
\affiliation{IFIN-HH, 077125 M\v{a}gurele-Bucharest,
Romania}

\author{Pawel Danielewicz}
\email{danielewicz@nscl.msu.edu}
\affiliation{National Superconducting Cyclotron Laboratory and
Department of Physics and Astronomy, Michigan State
University, East Lansing, Michigan 48824, USA}

\author{Che Ming Ko}
\affiliation{Cyclotron Institute and Department of Physics and
Astronomy, Texas A$\&$M University, College Station, Texas 77843,
USA}
\author{Akira Ono}
\email{ono@nucl.phys.tohoku.ac.jp}
\affiliation{Department of Physics, Tohoku University, Sendai
980-8578, Japan}

\author{Manyee Betty Tsang}
\email{tsang@nscl.msu.edu}
\affiliation{National Superconducting Cyclotron Laboratory and
Department of Physics and Astronomy, Michigan State
University, East Lansing, Michigan 48824, USA}

\author{Rui Wang}
\affiliation{Shanghai Institute of Applied Physics, Chinese Academy of Sciences, Shanghai $201800$, China}
\affiliation{Key Laboratory of Nuclear Physics and Ion-beam Application~(MOE), Institute of Modern Physics, Fudan University, Shanghai $200433$, China}
\author{Hermann Wolter}
\email{hermann.wolter@physik.uni-muenchen.de}
\affiliation{Physics Department, University of Munich, D-85748 Garching, Germany}

\author{Jun Xu}
\email{xujun@zjlab.org.cn}
\affiliation{Shanghai Advanced Research Institute, Chinese Academy of Sciences,
Shanghai 201210, China}
\affiliation{Shanghai Institute of Applied Physics, Chinese Academy
of Sciences, Shanghai 201800, China}

\author{Zhen Zhang}
\affiliation{Sino-French Institute of Nuclear Engineering $\&$
Technology, Sun Yat-sen University, Zhuhai 519082, China}

\author{Lie-Wen Chen}
\affiliation{
School of Physics and Astronomy, Shanghai Key Laboratory for Particle Physics and Cosmology, and Key Laboratory for Particle Astrophysics and Cosmology (MOE), Shanghai Jiao Tong University, Shanghai 200240, China}


\author{Hui-Gan Cheng} 
\affiliation{
School of Physics and Optoelectronic Technology, South China University of Technology, Guangzhou 510641, China 
}




\author{Hannah Elfner}
\affiliation{Frankfurt Institute for Advanced Studies, Johann Wolfgang Goethe University, Ruth-Moufang-Strasse 1, 60438 Frankfurt am Main, Germany}
\affiliation{Institute for Theoretical Physics, Goethe University, Max-von-Laue-Strasse 1, 60438 Frankfurt am Main, Germany}
\affiliation{GSI Helmholtzzentrum f\"{u}r Schwerionenforschung, Planckstr. 1, 64291 Darmstadt, Germany}
\author{Zhao-Qing Feng}
\affiliation{
School of Physics and Optoelectronic Technology, South China University of Technology, Guangzhou 510641, China} 
\author{Myungkuk Kim}
\affiliation{Department of Physics, Ulsan National Institute of Science and Technology, Ulsan 44919, Korea}
\author{Youngman Kim}
\affiliation{Rare Isotope Science Project, Institute for Basic Science, Daejeon 34000, Korea}
\author{Sangyong Jeon}
\affiliation{Department of Physics, McGill University, Montreal, Quebec 
H3A2T8, Canada}
\author{Chang-Hwan Lee}
\affiliation{Department of Physics, Pusan National University, Busan 46241, Korea}
\author{Bao-An Li}
\affiliation{Department of Physics and Astronomy, Texas A$\&$M
University-Commerce, Commerce, Texas 75429-3011, USA}

\author{Qing-Feng Li}
\affiliation{School of Science, Huzhou University, Huzhou 313000,
China}
\affiliation{Institute of Modern Physics, Chinese Academy of
Sciences, Lanzhou 730000, China}

\author{Zhu-Xia Li}
\affiliation{China Institute of Atomic Energy, Beijing 102413, China}

\author{Swagata Mallik}
\affiliation{Physics Group, Variable Energy Cyclotron Centre,
1/AF Bidhan Nagar, Kolkata 700064, India}




\author{Dmytro Oliinychenko}
\affiliation{Lawrence Berkeley National Laboratory, 1 Cyclotron Rd., Berkeley, California, USA, 94720}    
\affiliation{Institute for Nuclear Theory, University of Washington, Seattle, WA, 98195, USA}     


\author{Jun Su}
\affiliation{Sino-French Institute of Nuclear Engineering $\&$
Technology, Sun Yat-sen University, Zhuhai 519082, China}

\author{Taesoo Song}
\affiliation{Frankfurt Institute for Advanced Studies, Johann Wolfgang Goethe University,
Ruth-Moufang-Strasse 1, 60438 Frankfurt am Main, Germany}
\affiliation{Institut f\"{u}r Theoretische Physik, Universit\"{a}t Gie\ss en, Heinrich-Buff-Ring, 35392 Gie\ss en, Germany}

\author{Agnieszka Sorensen}
\affiliation{Department of Physics and Astronomy, University of California, Los Angeles, CA 90095, US}


\author{Feng-Shou Zhang}
\affiliation{Key Laboratory of Beam Technology and Material
Modification of Ministry of Education, College of Nuclear Science
and Technology, Beijing Normal University, Beijing 100875, China}
\affiliation{Beijing Radiation Center, 100875 Beijing, China}


\begin{abstract}
Within the transport model evaluation project (TMEP) of simulations for heavy-ion collisions, the mean-field response is examined here.  
Specifically, zero-sound propagation is considered for neutron-proton symmetric matter enclosed in a periodic box, at zero temperature and around normal density.  The results of several transport codes belonging to two families (BUU-like and QMD-like) are compared among each other and to exact calculations. For BUU-like codes, employing the test particle method, the results depend on the combination of the number of test particles and the spread of the profile functions that weight integration over space. These parameters can be properly adapted to give a good reproduction of the analytical zero-sound features. QMD-like codes, using molecular dynamics methods, are characterized by large damping effects, attributable to the fluctuations inherent in their phase-space representation. Moreover, for a given nuclear effective interaction, they generally lead to slower density oscillations,  
as compared to BUU-like codes. The latter problem is mitigated in the more recent lattice formulation of some of the QMD codes. The significance of these results for the description of real heavy-ion collisions is discussed. 
\end{abstract}


\maketitle

\section{Introduction}
\label{sec:Intro}

A large variety of phenomena, ranging from the structure of nuclei and their
decay modes up to the life and the properties of massive stars, are governed by the nuclear Equation of State (EoS), thus giving great importance to dedicated studies. In particular, the understanding of the properties of exotic nuclei, as well as neutron stars and supernova dynamics, entails the knowledge of
the behavior of nuclear symmetry energy, on which several investigations are concentrating nowadays 
\cite{science06,Baran_PR410_05,fuchs06,Steiner_PR411_05,BALi_PR464_08,Horow_way-forw_14,SymmE_EPJA50_14,Oertel2017,colonna20}.

In the laboratory, heavy-ion collisions are the primary way to investigate nuclear matter away from saturation conditions. States of high density and excitation can be created on short time scales. However, these are complex non-equilibrium processes. The challenge is to connect nuclear matter states of interest to the final observables, so that information on the EoS can be extracted. Transport approaches are the main tool to extract this information. Therefore, the reliability of transport studies of heavy-ion collisions and the robustness 
of their predictions is 
important in heavy-ion research.



It has recently become apparent that different conclusions could be drawn from the same data by relying on transport simulations, e.g.,\ in the investigations of isospin equilibration in peripheral collisions (isospin diffusion) 
\cite{Tsang04,Galichet09, Tsang09, Rizzo08, Bao_1,Bao_2}, 
or in the interpretation of ratios of charged pions \cite{Xiao09, Cozma17, Hong14, Xie13, ZQFeng10, Song15}.
These discrepancies could naturally derive from the different approximation schemes, adopted in the different transport models, to deal with the quantum many-body problem
or from differences in various technical assumptions. Indeed, because of the complexity of transport equations, and in particular of their dimensionality, 
they are solved by simulations, which requires the use of sophisticated algorithms that invoke statistical sampling and finite phase-space resolutions. The impacts of these numerical details on predictions and conclusions are often difficult to discern.  This situation led to the idea of a systematic comparison and evaluation of transport codes under controlled conditions, to eventually provide benchmark calculations and thus to improve the ability to reach robust conclusions from the comparison of transport simulations with experimental data. 

Previous studies along this direction were dedicated to the comparison of transport model predictions for Au + Au collisions \cite{Kolomeitsev05,Xu2016}.
The compared aspects mainly included the stability of the initialized nuclei, the effectiveness of Pauli blocking for the final states of nucleon-nucleon (N-N) collisions, and  predicted flow observables.  
There were indications that a large part of the observed differences in the predicted reaction path and corresponding observables (such as collective flows) resulted from differences in the initialization of the systems and in the treatment of the collision integral (mainly Pauli blocking effects). The mean-field dynamics also seemed to play a role.
However, the origins of the differences were often difficult to pin down unambiguously, since various effects interplay and propagate.

Significant progress in understanding the behavior of the different transport codes was made with subsequent studies, based on box calculations, i.e.,~simulations of nuclear matter enclosed in a box with imposed periodic boundary conditions. In particular, the box calculations have the advantage that 
the different aspects of heavy-ion collisions can be isolated and tested separately, e.g., the description of N-N scattering processes (i.e., two-body correlations) and the  mean-field dynamics. Whereas features of the collision integral, such as Pauli blocking effects and meson (pion) production, have been the object of our recent studies \cite{comp2,comp3},  the investigation of the mean-field dynamics is the aim of the present paper.

To test the mean-field dynamics in a box in this work, we investigate a typical example of collective motion, namely the zero-sound propagation, i.e. 
the mean-field propagation of a disturbance of the single-particle distribution
in nuclear matter. We initialize a disturbance by setting up a standing wave in density, and by assigning the momenta of the particles randomly in the local Fermi sphere, as commonly done in transport codes. 
This wave is then propagated by the Vlasov part of the different transport models using density 
functionals that give identical EoS features, and the corresponding results are compared with
each other. 
This will allow to see characteristic differences between the different types of transport codes, as well as the dependence on calculational parameters.

One should notice that, for box calculations, there are in some cases exact limits available from kinetic theory or Landau theory, against which the performance of the codes can be judged, instead of against each other.

However, in comparing the different codes against each other and against any known limits, one should keep in mind that: (1) there are different families of transport theories: 
Boltzmann-Vlasov-type codes (usually referred
to under the name of Boltzmann-Uehling-Uhlenbeck (BUU)) and molecular dynamics-type codes (usually Quantum Molecular Dynamics (QMD)).  
The two families of codes start from different theoretical frameworks and/or different philosophies in modeling heavy-ion collisions. Thus, one cannot expect that they completely agree with each other; (2)  
basic differences may be present between exact limits from kinetic theory and simulations, implying that the exact limit cannot actually be reached. These may lie, e.g.,  
in unavoidable fluctuations in a concrete simulation strategy.
The effects of numerical fluctuations were already explored in Ref.\cite{comp2}.  
However, 
differences between codes of the same type and differences with the exact limits in many cases can suggest improvements of the codes. 

While the zero-sound motion is here a specific example for our investigation of transport codes, it is by itself an interesting phenomenon, which we are able to study in detail. In the limit of small amplitudes, exact results for the frequency can be derived from Landau theory, where relativistic effects, or more generally effects of the effective mass, can be studied. 
We also note that mean-field studies have been devoted in the past to investigate collective motion in finite nuclei, both with semi-classical transport theories as here and with time-dependent Hartree-Fock (TDHF) theory
\cite{Zheng2016,Kong2017,roca,Burrello2019,Maruhn}.

Since small amplitudes are not typical for a numerical study appropriate to heavy ion collisions, we use a large amplitude of the initial perturbation. This then leads to non-linear effects due to the non-linear terms in the force and to mode-mixing. 
Furthermore, the damping of the wave is an important question, which here is not only due to Landau damping, i.e. mode mixing, but also due to fluctuations
that may arise from the numerical resolution of the phase space.
Thus, the mean-field analysis presented in this work can be considered as a valuable test also 
for the general case of the mean-field dynamics involved in heavy-ion collisions at intermediate
energies, which is largely influenced by the emergence of collective phenomena.   

For the simulations presented in this work, we employed the same main protocol 
as developed in the context of Refs.\cite{Xu2016,comp2,comp3}. Contributors of the participating codes performed specified ``homework" calculations. The resulting files were sent to the writing group 
for evaluation and preparation of publication. The results were then discussed in several 
meetings (see, for instance, the NuSym series of conferences, and in particular
in Ref.\cite{NuSym2018}).   

The article is organized as follows: a short description of the two families of transport approaches is given in Sect.~\ref{sec:transport_approaches}, to state the main differences between the approaches and clarify the terminology.   The homework specifications pertaining to this paper are described in Sect.~\ref{sec:homework_description}. 
Analytical and reference results relating to the present comparison are 
presented in Sects.
\ref{sec:Analytic} and \ref{sec:DFS}.
The results of the comparison are described in following three srctions: In 
Sect. \ref{sec:oscill}, we discuss the coordinate space evolution, and questions of the global momentum and energy distributions. We then explore the evolution in wave number and frequency space via spatial and temporal Fourier transforms, for selected codes in each of the two families in 
Sect. \ref{sec:selected}, and for all codes in Sect. \ref{sec:all}.
Finally, a discussion of the results, conclusions and an outlook
can be found in Sect. \ref{sec:Discussion}. 

The participating codes and their contributors are listed in Table \ref{tab:codes}. 
The major codes used presently in the interpretation of heavy-ion collisions are represented, with 
nine of the BUU-type and five of the QMD-type. 
The codes can be classified according to their treatment of relativity: non-relativistic codes, codes with relativistic kinematics, and codes with relativistic dynamics in a relativistic mean-field (RMF) formulation (labelled "cov" in Table \ref{tab:codes}).
We note that the well-known antisymmetrized molecular dynamics (AMD) code \cite{AMD} is not included in the present comparison, since a box condition in this code is not comparable to the treatment in the semi-classical codes.

\begin{table*}[htbp]
\newcommand{\tabincell}[2]{\begin{tabular}{@{}#1@{}}#2\end{tabular}}
\caption{\label{tab:codes} The acronyms, code correspondents for the calculations shown in this paper, 
dynamical treatment (nonrelativistic/relativistic kinematics/covariant), (test) particle features 
and representative references of the nine BUU-type and five QMD-type codes
participating in the present comparison.}
    \begin{tabular}{ccccccc}
    \hline
    \hline
    Type & Acronym & Code Correspondents & Rel/Non-Rel & Particle profiles &
\tabincell{c}{$(\Delta x)^2$ [fm$^2$]\footnote{$\Delta x$ is the width of the Gaussian wavepacket as in Eq.(\ref{eq:QMDwf}).}\\ or $l$ [fm]\footnote{$l$ is the half base 
for test particles with triangular or trapezoid profile. See 
Refs.\cite{Lenk89} and \cite{Pawel} for more details.}} & Reference \\
    \hline
BUU  &    BUU-VM\footnote{BUU code developted jointly at VECC and McGill.}& S.~Mallik
     & non-rel &  triangle & 1 & \cite{VECC}\\
   &  DJBUU  & Y.~Kim & cov & $[1 - (|\vec{r}|/\Delta x)^2]^3$ & 6.25 & \cite{Kim} \\
   &  GiBUU & J.~Weil & cov & Gaussian & 1 & \cite{Gaitanos} \\
   &  IBUU\footnote{There is also a new version of this code (IBUU-L) in the comparison \cite{ref_IBUUL}, explained in Sect. \ref{lattice}.}{}   &  J.~Xu & rel & triangle & 1 & \cite{BALi08} \\
   &  LHV & R.~Wang & rel & triangle & 2 & \cite{Wang19} \\
  &  pBUU   &  P.~Danielewicz & cov & trapezoid & 0.92 & \cite{Pawel}\\
   &  RVUU   & Z.~Zhang & cov & point & 0 & \cite{Song15,KoCM}\\
   &  SMASH & A.~Sorensen & cov & triangle & 2 & \cite{Weil}\\
   &  SMF    & M.~Colonna & non-rel & triangle & 2 & \cite{Colonna} \\
     \hline
QMD   &  ImQMD\footnote{ImQMD-CIAE in Ref.\cite{Xu2016}. There exists also a Lattice version of the code, ImQMD-L \cite{Limqmd}, see Sect. \ref{lattice}. } & Y.~X.~Zhang & rel & Gaussian & 2 & \cite{YXZhang}\\
   &  IQMD-BNU & J.~ Su & rel & Gaussian & 2 & \cite{JunS}\\
   &  IQMD-IMP\footnote{Also known as LQMD in literature.} & Z.~Q.~Feng & rel 
& Gaussian & 2 &\cite{ZQFeng}\\
   &  TuQMD & D.~Cozma & rel & Gaussian & 2 & \cite{Cozma}\\
   &  UrQMD & Y.~J.~Wang & rel & Gaussian & 2 & \cite{QFLi,Bass} \\
    \hline
    \hline
    \end{tabular}
\end{table*}

\section{Transport approaches}
\label{sec:transport_approaches}


The primary methodology for
the dynamics of nuclear collisions at Fermi/intermediate energy
are 
semi-classical transport theories, such as the Nordheim approach, in which the Vlasov
equation for the one-body phase space distribution, $f(\vec{r},\vec{p}; t)$, is extended with a Pauli-blocked
Boltzmann collision term \cite{ber88,bon94}, 
which accounts for the average effect of the two-body residual
interaction. The thus resulting transport equation, often called Boltzmann-Uehling-Uhlenbeck (BUU)
equation, contains two main ingredients: the self-consistent mean-field potential and the two-body
scattering cross sections. In order to introduce fluctuations and further (many-body) correlations in the treatment of the reaction dynamics, a number of different avenues have been undertaken, which can be differentiated into
two classes (see Refs.\cite{xu19,colonna20,onoPPNP}
 for recent reviews).
One is the class of molecular
dynamics (MD) models \cite{ono99,aic91,fel90,col98,ono92,pap01,zha06},
 while the other kind is represented by 
stochastic extensions of mean-field approaches of the BUU type 
\cite{ayi88,abe96,ran90,cho04,Napoli}. 

\subsection{{BUU-like models}}
In BUU-like approaches, the time evolution of the one-body
phase space distribution function, $f(\vec{r},\vec{p};t)$,  follows the 
equation
\begin{equation}
\Big(\frac{\partial}{\partial t}+ \vec{\nabla}_p\epsilon \cdot \vec{\nabla}_r-\vec{\nabla}_r \epsilon\cdot \vec{\nabla}_p\Big) f(\vec{r},\vec{p};t)=I_{\text{coll}} (\vec{r},\vec{p};t) \, ,
\label{eq:BUU}
\end{equation}
where $\epsilon[f]$ is the single-particle energy, usually derived from a density functional, and $I_{\text{coll}}$ is the two-body collision integral, specified by an in-medium cross section $d\sigma^\text{med}/d\Omega$. 
Fluctuations of the one-particle density,
which should account for the effect of the neglected many-body correlations, can be introduced by adding
to the r.h.s. of Eq.(\ref{eq:BUU})
 a stochastic term, representing the fluctuating part of the collision integral \cite{ayi88,abe96,ran90}. This leads to the Boltzmann-Langevin (BL) equation, in close analogy with the Langevin equation for a Brownian motion. 

In the present study, we focus on the mean-field propagation, thus we neglect
the r.h.s. of Eq.(\ref{eq:BUU}) and any fluctuation terms. 
It should be noticed that the BUU theory can 
more generally be formulated in a relativistic framework, and actually most codes in this comparison use a relativistic formulation. 
{
In the relativistic covariant approach, the nucleons are coupled to 
momentum-independent scalar and vector fields. 

Let us introduce the kinetic momentum ${p^*}^\mu = p^\mu - A^\mu$ and the energy 
$E^*\equiv {p^*}^0 = \sqrt{\vec{p^*}^2 + {m^*}^2}$. Here $A^\mu$ represents the vector field;
the Dirac mass, $m^*$, is given by $m^* = M - \Phi$, with $\Phi$ denoting the 
scalar field and $M$ the nucleon mass. 
The vector field depends on the baryon four-current $j^{\mu}(\vec{r};t)$, which, in the local density approximation, is
given self-consistently by: 
\begin{equation}
\vec{j} = 4\int \frac{d^3p}{(2\pi)^3} ~ \frac{\vec{p^*}}{E^*}~ f(\vec{r},\vec{p};t)
\label{eq:current}
\end{equation}
and
\begin{equation}
j^{0} \equiv \rho = 4\int \frac{d^3p}{(2\pi)^3} ~ f(\vec{r},\vec{p};t)~,
\label{eq:density}
\end{equation}
where $\rho(\vec{r};t)$ is the nucleon density
{ and the factor $4$ is due to the spin and isospin degeneracies
of nucleons in symmetric nuclear matter considered here.} 
Similarly, the scalar field $\Phi$ depends on the scalar density $\rho_S(\vec{r};t)$, which is defined as: 
\begin{equation}
{\rho_S} = 4\int \frac{d^3p}{(2\pi)^3} ~ \frac{{m^*}}{E^*}~ f(\vec{r},\vec{p};t).
\label{eq:scalar}
\end{equation}
The single-particle energy 
in Eq.(\ref{eq:BUU}) simply reads:
$\epsilon = p^0 = E^* + A^0$. The specific dependence of the fields on the densities is detailed in Sect.\ref{ingr_cov}. 

It is of interest to introduce the energy density, $e(\rho,T)$, for nuclear
matter at rest, from which the 
nuclear matter EoS at the temperature $T$ is directly derived. 
Considering that the current $\vec{j}$ vanishes, $e(\rho)$ can be expressed as: 
$$
e(\rho,T) =  4\int \frac{d^3p}{(2\pi)^3} ~  \sqrt{\vec{p}^2 + {m^*}^2}~ f_{FD}(p,T)
+ \int_0^{\rho_S} d\rho'_S \rho'_S \frac{d\Phi}{d\rho'_S}~   
$$
\begin{equation}
+~\int_0^{\rho} d\rho'A^0(\rho').~~~~~~~~~~~~~~~~~~~~~~~~~~~~~ 
\label{eq:en_den}
\end{equation}
Here, the function $f_{FD}(p,T)$ 
denotes the local Fermi-Dirac distribution at the 
temperature considered. 

As was mentioned, 
the transport codes that we consider in the following can be assigned
to three main categories: 

(a) {\it Non-relativistic codes} (labelled as ``non-rel'' in Table I). 

These codes can be framed into the general scheme illustrated above, 
if one considers only vector fields, and neglects the spatial components 
of the baryon four-current ($\vec{j}$ = 0). Thus the energy $E^*$ becomes 
$E^* \rightarrow~ E = \sqrt{\vec{p}^2 + M^2}$. Moreover, in this case, the non-relativistic
limit is taken for $E$. Thus the single-particle energy can be written as:
$\epsilon = \frac{\vec{p}^2}{2M} + U(\rho) + M $, where $U(\rho)$ is the mean-field potential, {which is introduced phenomenologically}. 
A simple Skyrme-like form will be employed here: $U(\rho) = a(\rho/\rho_0) + b(\rho/\rho_0)^\sigma$,
where $\rho_0$ denotes the saturation density {and the non-linear
term takes into account the effect of many-body forces}.

(b) {\it Codes with relativistic kinematics} (labelled as ``rel'' in Table I).

The same ingredients as in the ``non-rel'' case are considered, but in this case the kinematics is relativistic. Hence, the single-particle energy is expressed
as: $\epsilon = E + U(\rho)$.

(c) {\it Covariant codes} (labelled as ``cov'' in Table I).

We place into this category all codes that employ scalar fields and/or vector
fields depending on the baryon four-current $j^\mu$. 
} 

\subsubsection{{Ingredients of the covariant codes}}
\label{ingr_cov}
{ In this section, we give more details about the codes of the latter
category, namely the codes labelled as ``cov'' in Table I. 

\noindent
RVUU: This code follows the scheme of the standard (non-linear) Walecka model.  
Denoting by $m_\sigma, m_\omega$ and $g_\sigma, g_\omega$ the masses and coupling
constants of the $\sigma$ (scalar) and $\omega$ (vector) mesons, respectively, 
the following relations hold for scalar and vector fields: 
\begin{equation}
{\rho_S} = \frac{m_\sigma^2}{g_\sigma^2}\Phi + \frac{A}{g_\sigma^3}\Phi^2
+ \frac{B}{g_\sigma^4}\Phi^3~~~;~~~ A^\mu = \frac{g_\omega^2}{m_\omega^2}j^\mu.
\end{equation} 
The corresponding energy density, for nuclear matter at rest, 
is (see Eq.(\ref{eq:en_den})): 
$$
e(\rho,T) =  4\int \frac{d^3p}{(2\pi)^3} ~ \sqrt{\vec{p}^2 + {m^*}^2} 
f_{FD}(p,T) +
\frac{m_\sigma^2}{2~g_\sigma^2}\Phi^2 + 
$$
\begin{equation}
+\frac{A}{3~g_\sigma^3}\Phi^3 + \frac{B}{4~g_\sigma^4}\Phi^4 + \frac{g_\omega^2}{2~m_\omega^2}\rho^2.  
\end{equation}
\noindent
DJBUU: This code adopts the approximation of neglecting the spatial 
components of the baryon four-current ($\vec{j}$ = 0), so that the single-particle energy is 
given by $\epsilon = \sqrt{\vec{p}^2 + {m^*}^2} + A^0$, 
whereas the nuclear matter energy density keeps the same expression as
in RVUU.

\noindent
pBUU: In the version of the pBUU model employed for the homework, only a scalar
field is considered, so that the single-particle energy simply reads:
$\epsilon = \sqrt{\vec{p}^2 + {m^*}^2}$. The scalar field $\Phi$ is defined as:
\begin{equation}
-\Phi(\rho_S) \equiv U(\rho_S) = \frac{a(\rho_S/\rho_0) + b(\rho_S/\rho_0)^\sigma}{1 +
(\frac{\rho_S/\rho_0}{2.5})^{\sigma-1}}
\label{eq:pBUU_field}
\end{equation}
The role of the denomimator in Eq.(\ref{eq:pBUU_field}) is to prevent 
supraluminous behavior at high densities. The energy density is 
calculated from Eq.(\ref{eq:en_den}). 
We notice that the scalar field adopted here is quite close to the Skyrme
parametrization used for the mean-field potential of categories ``non-rel'' 
and ``rel''.  

\noindent
SMASH: In the SMASH code, no scalar field is considered, but a more complex
vector field, $A^\mu = \sum_i A_i^\mu(\vec{r};t) = \sum_i C_i (j_\nu j^\nu)^{\frac{\beta_i}{2}-1}~j^\mu$, is introduced, 
leading to an overall attractive potential.  
Thus the single-particle energy is given as: 
$\epsilon =  \sqrt{\vec{p^*}^2 + M^2} + \sum_i C_i (j_\nu j^\nu)^{\frac{\beta_i}{2}-1}~\rho$.   
For nuclear matter at rest, the corresponding energy density reads: 
\begin{equation}
e(\rho,T) =  4\int \frac{d^3p}{(2\pi)^3} ~  \sqrt{\vec{p}^2 + M^2} f_{FD}(p,T) +
\sum_i \frac{C_i}{\beta_i}\rho^{\beta_i}.
\end{equation}
We note that, contrary to SMASH, in RVUU and DJBUU 
the linear vector field is repulsive (as in the standard Walecka model), whereas the scalar field leads to an attractive potential.}

\subsubsection{{Numerical solution of the transport equations}}
The integro-differential non-linear BUU equation is solved numerically. To this end, the distribution function is represented in terms of finite elements, so-called test particles (TP)~\cite{Wong82}, as
\begin{equation}
f(\vec{r},\vec{p};t)=\frac{(2\pi)^3}{4 N_{\text{TP}}} \sum_{i=1}^{AN_\text{TP}} G(\vec{r}-\vec{R}_i(t)) \, \tilde{G}(\vec{p}-\vec{P}_i(t)) \, ,
\label{eq:fTP}
\end{equation}
where $N_{\text{TP}}$ is the number of TP per nucleon
(set to 100 in this work),
$\vec{R}_i$ and $\vec{P}_i$  are the time-dependent centroid coordinates and momenta of the TPs, and $G$ and $\tilde{G}$ are the profile functions in coordinate and momentum space, respectively, with a unit norm (e.g.\ $\delta$ functions or normalized Gaussians). In particular, $\delta$ functions are generally adopted
in momentum space.  
We remind the reader that the degeneracy factor $4$ (in the
denominator of Eq.(\ref{eq:fTP})) is to define $f(\vec{r},\vec{p},t)$ as the spin-isospin averaged phase space occupation probability, which is well suited to 
the case considered here (symmetric matter). 
It is also possible to express the distribution function for each isospin (or spin) state in a similar way.  Upon inserting the ansatz Eq.~\eqref{eq:fTP} into the left-hand side of Eq.~\eqref{eq:BUU}, i.e., without the collision integral, Hamiltonian equations of motion for the TP centroid propagation follow:
\begin{equation}
\frac{d\vec{R}_i}{dt}=\vec{\nabla}_{{P}_i} \epsilon \hspace*{2em} \text{and} \hspace*{2em} \frac{d\vec{P}_i}{dt}=-\vec{\nabla}_{R_i} \epsilon \, .
\label{eq:prop}
\end{equation}
{The treatment of the collision integral is discussed in detail
in Ref.\cite{comp2}, but this is not of relevance in the present study of only 
Vlasov dynamics.} 

\subsection{{QMD models}}
In quantum molecular dynamics (QMD) models, the many-body state is represented by a simple product wave function of single-particle states
with or without antisymmetrization \cite{ono99,aic91}. 
The single-particle wave functions are usually assumed to have
a fixed Gaussian shape. In this way, though the nucleon wave functions are 
independent
(mean-field approximation), the use of localised wave packets induces classical many-body correlations both in
the mean-field propagation and two-body in-medium scattering (collision integral), where the latter is treated stochastically.
Hence, this way to introduce many-body correlations and produce a possible trajectory branching is
essentially based on the use of localized nucleon wave packets.
{It has been proven to be particularly efficient for the description of fragmentation events, where nucleons are well localized inside separate fragments in the final state \cite{aic91}.}
The time evolution of nuclear dynamics is formulated in terms of the changes in nucleon coordinates and momenta, similar to classical molecular dynamics, which are the centroids of the wave packets.
They move under the influence of nucleon-nucleon interactions, which are usually consistently accounted for by density functionals. The method can also be viewed as derived from the time-dependent Hartree method with a product trial wave function of single-particle states in Gaussian form
\begin{align}\label{eq:QMDwf}
&\Psi(\vec{r}_1,\dots, \vec{r}_A; t) =  \prod_{i=1}^A \phi_i(\vec{r}_i;t), \\
&\phi_i(\vec{r}_i;t) = \frac{1}{[2\pi (\Delta x)^2\big]^{\frac{3}{4}}}\exp\bigg[-\frac{[\vec{r}_i-\vec{R}_i(t)]^2}{4(\Delta x)^2}\biggr]e^{(i/\hbar)\vec{P}_i(t)\cdot\vec{r}_i}. \nonumber
\end{align}
The centroid positions $\vec{R}_i(t)$ and momenta $\vec{P}_i (t)$ are treated as variational parameters within the variational principle for the time-dependent Hartree equation.
The widths $\Delta x$ are kept fixed and thus are not variational parameters, in order for the wave function to be able to describe finite distance structures, as observed in the fragmentation of colliding nuclei.
This strategy yields equations of motion for the coordinates of the wave packets of similar form as obtained for the TPs in BUU. 
{ The QMD codes that we will consider here employ relativistic
kinematics, thus they fall into the ``rel'' category.} 
This method has been extended to include anti-symmetrization in the wave function in the AMD method \cite{AMD}, which makes the equations of motion more complicated but with similar principles.

The main difference between the two methods lies in the amount of fluctuations and correlations in the representation of the phase space distribution. In the standard BUU approach, the phase space distribution function is seen as a one-body quantity and a smooth function of coordinates and momenta, which can be approximated increasingly better by increasing the number of TPs in the representation. In the limit of $N_{\text{TP}}\rightarrow \infty$, the BUU equation is solved exactly. In this limit the solution is deterministic and does not contain fluctuations.  However, as mentioned above, if fluctuations are considered to be important, 
suitable stochastic extensions can be formulated. 
Of course, numerical fluctuations are present in practical BUU calculations with a finite number of TPs.

In QMD, 
{nucleon correlations arise from}
the representation in terms of a finite number of wave packets of finite 
width, {leading to enhanced fluctuations of the one-body density.}
Thus, in the philosophy of QMD one wants to go beyond the mean-field approach and include correlations and fluctuations from the beginning. However, these fluctuations, which are essentially of classical nature, can lead to a loss of the fermionic character of the system more rapidly {than in BUU, as it was 
studied in Ref.\cite{comp2}}. 
The fluctuations in QMD-type codes are regulated and smoothed by choosing the parameter $\Delta x$, the width of the wave packet, cf. Eq.~\eqref{eq:QMDwf}.  
QMD can be seen as an event generator, where the time evolution of different events is solved independently and therefore the fluctuations among events are not suppressed even in the limit of an infinite number of events. 

The effects of this difference in the amount of fluctuations between the two approaches will clearly be seen in the comparisons that will follow.

\subsection{Lattice Hamiltonian and particle propagation}
\label{lattice}
{The solution of the (test) particle equations of motion, 
Eq.(\ref{eq:prop}), requires the calculation of the local single-particle 
energy, which also depends on the local density $\rho(R_i,t)$.
The latter can be evaluated starting from Eq.(\ref{eq:fTP}), with $N_{TP}$ = 1
in the QMD case.
}
Some of the codes (of type ``non-rel'' or ``rel'') involved in our comparison employ the Lattice Hamiltonian
framework \cite{Lenk89}. 
This method has been proven to be particularly effective for the numerical 
solution of the Vlasov equation, especially as far as energy conservation is concerned. 
Namely, the coordinate space is divided into cubic cells
(typically of volume $\Delta l^3$ =  1 fm$^3$) and the spatial density is evaluated
at each cell site {coordinates}, $\vec{r_\alpha}$, {and given as 
$\rho_\alpha = \rho(\vec{r_\alpha})$}. 
Then, the potential part of the total Hamiltonian of the system is 
written as 
\begin{equation}
H_{pot}=\Delta l^3\sum_{\alpha}e_{pot}(\rho_\alpha),
\end{equation}
where $e_{pot}$ denotes the potential part of the energy density
and $\rho_\alpha = \rho(\vec{r_\alpha})$. 
{We remind that the density $\rho_\alpha$, and thus the Hamiltonian $H_{pot}$, depend on the (test) particle centroids, $\vec{R}_i(t)$, according to Eq.(\ref{eq:fTP})}. 
{With 
$\vec{P}_i(t)$ representing the momentum of the  $i^{th}$ test particle, 
the equation of motion from the Hamiltonian is then} 
\begin{equation}
 \frac{d\vec{P}_{i}}{dt} = 
-\Delta l^3\sum_{\alpha}\frac{de_{pot}}{d\rho_\alpha}
\vec{\nabla}_{{R}_i}G_\alpha= 
-\Delta l^3\sum_{\alpha}\epsilon_{pot}(\rho_\alpha)
\vec{\nabla}_{{R}_i}G_\alpha,
\label{pt}
\end{equation}
{where $\epsilon_{pot}$ denotes the potential part of the single-particle energy and $G_\alpha = G(\vec{r_\alpha} - \vec{R}_i)$.}
The Lattice Hamiltonian framework is adopted in BUU-VM, SMF, LHV, and in the 
Lattice version of IBUU (IBUU-L). 
In IBUU-L, a triangle profile function with $l$ = 2 fm is used for the test particles \cite{ref_IBUUL}.  
For ImQMD, we will also consider a Lattice version (ImQMD-L) that employs 
{a mesh with non-regular intervals},
 better suited to deal with Gaussian particle profile 
functions \cite{Limqmd}.

\section{Homework description}
\label{sec:homework_description}


The understanding of mean-field effects is essential to reach a reliable description of the dynamics of nuclear reactions. A dedicated homework has been devised to test the mean-field propagation under controlled situations in the different transport codes.
To that purpose, we consider uniform nuclear matter at zero temperature in a box with periodic boundary conditions. The system is perturbed by building up
the density profile along one direction (the z axis, in our case). The initial density perturbation is then propagated by motion in the nuclear mean-field and the time evolution of the system is followed until the time $t_{fin}$.  The collision integral is turned off and rather simple mean-field parametrizations are adopted, giving the correct saturation properties and a selected value of the compressibility modulus, $K_0$. This then corresponds to a pure Vlasov mode for the transport codes. Thus BUU-like trajectories should be fully
deterministic (apart from numerical fluctuations), whereas in the QMD case
the presence of fluctuations is intrinsic to the model. 

As already noticed for the box comparisons involving the collision integral \cite{comp2,comp3}, differences are expected in the results of QMD-like  and BUU-like codes, mainly due to the larger amount of fluctuations and the larger width of the particle wave packet employed
in QMD codes.  Indeed, fluctuations influence the damping of the density oscillations, whereas the packet width affects the calculations of the mean-field potential and thus the oscillation frequency.  

The goal of the homework is to understand the propagation of initial 
sinusoidal perturbations by the nuclear mean-field, and thus to check the dispersion relation for the mean-field propagation of density fluctuations 
(zero-sound propagation).  Thus, the average density in $z$-direction at different times, $\langle\rho(z,t)\rangle$, represents one of the main quantities to be extracted from the calculations and analysed.  The calculations are averaged over many events to try to understand 
the average mean-field behavior, which is the quantity of interest in a heavy-ion collision.
A rather compact and effective representation of the behavior of the system  is provided by the time evolution of the  spatial Fourier transform, $\rho_{k}(t)$. Additional insight can be obtained by a further Fourier transform in time, leading to the response function $\rho_k(\omega)$.

The box calculations are performed with periodic boundary conditions \cite{comp2,comp3}. Reflecting boundary conditions are not used because they could give rise to edge effects, negligible only in the limit of very large boxes.
In contrast, with periodic boundary conditions the box can be kept relatively small with no significant finite-size effects. The dimensions of the cubic box are $L_\alpha = 20 \, \text{fm}$, $\alpha \equiv x,y,z$. The position of the center of box is ($L_x$/2, $L_y$/2, $L_z$/2).
In a periodic box, a particle that leaves the box on one side should enter it from the opposite side with the same momentum.
Once a coordinate $\alpha$ ventures outside of the box, it may be reset with $r_\alpha \rightarrow \mathop{\text{modulo}}( r_\alpha,  L_\alpha )$. Similarly, the separation between two points $\Delta r_{ij,\alpha}=r_{i,\alpha}-r_{j,\alpha}$ must be redefined as $\Delta r_{ij,\alpha} \rightarrow \mathop{\text{modulo}}( \Delta r_{ij,\alpha}+L_\alpha/2, L_\alpha ) - L_\alpha/2$.
This method is completely sufficient and will cope with all structures, as long as the characteristic lengths are short relative to $L_\alpha$/2. 

This periodic box condition applies only to classical or semiclassical approaches.
In quantum mechanical approaches such as in AMD \cite{AMD}, the implementation of a periodic box calculation is more involved, since now the wave functions have to satisfy the boundary condition, implying that the momenta become discretized in steps of the order of $\Delta p= 2\pi/L_\alpha\approx$ 62 MeV/c, which is not so much smaller than the Fermi momentum. A special code would have to be written for this, which would not be comparable to the semi-classical codes, and would also be very different from the code used for heavy-ion collisions. However, in this box comparison we want to change the codes as little as possible from those used for heavy-ion collisions.
Thus results from the AMD code are not included in this comparison.

\subsection{Details of the homework}
\label{sec:hw_details}

We consider symmetric nuclear matter at saturation
density $\rho_0$ = 0.16~fm$^{-3}$ and zero temperature. 
For the cubic box employed (of size $L_\alpha$  = 20 fm),
this corresponds to $A=1280$ nucleons.
The simulations are followed until $t_{fin}$= 500 fm/c, with a recommended time 
step of either $\Delta t$=0.5 or 1.0 fm/c. 
A detailed description of the homework is as follows. 
\subsubsection{Initialization} 
The system is initialized by impressing a 
sinusoidal distortion with wave number $k$ and amplitude $a_\rho$ on the 
density in the box, along the $z$ direction:   
$\rho(z,t=0) = \rho_0  + a_\rho \sin(kz)$. 
It should be noticed that for QMD-type models, which employ Gaussian 
functions of sizeable width for the nucleon wave packets (such as
$G(\vec{r}) \approx \exp[-(\vec{r}-\vec{R_i})^2/(2(\Delta x)^2)]$, see Eq.(\ref{eq:QMDwf})), 
the specified density distribution can be obtained by sampling the centroids $\vec{R_i}$ of the wave packets according to the following density distribution:
$\rho(z,t=0)^{MD}=\rho_0+ a_\rho \exp[((\Delta x)^2 k^2)/2]\sin(kz)$ \cite{Cozma_priv}.
Because of the periodic boundary conditions imposed to the system, 
the wave number $k$ may take the values $k=n\,2\pi/L_\alpha$, with   
$n =1,...,L_\alpha/(2~\Delta z)$, where $\Delta z$ is the spatial 
step along the z direction.
In the homework, we will test small wave numbers ($n=1$), which are less affected by the surface effects induced by the finite width of 
the particle profile functions,  
and we adopt $a_\rho  = 0.2~\rho_0$.
We note that the amplitude is not small relative to the non-linearities of the 
mean-field, so the sinusoidal wave is therefore distorted already after a short time of about 10 fm/c, as seen later.
The particle momenta are initialized randomly in a local Fermi sphere, with 
the Fermi momentum defined as a function of the local density of the
initialized density profile. 
\subsubsection{The nuclear interaction}
\label{nucl_int}
The Coulomb interaction and the nuclear symmetry force are turned off. The following simplified isoscalar nuclear force is employed:
For the codes of type "rel" and "non-rel"  
a standard Skyrme parametrization (without momentum-dependence) for the
single-particle potential is used,
\begin{equation}
U(\rho)  = a (\rho/\rho_0) + b (\rho/\rho_0)^\sigma,
\end{equation}
with the following parameters: $a = -105.716$ MeV, $b = 52.836$ MeV, $\sigma$ = 2.587. 
The nucleon mass is taken to be $M$ = 938 MeV.  
This parameterization leads to the following nuclear matter 
properties: compressibility $K_0=500$ MeV, 
saturation density $\rho_0 = 0.16~$fm$^{-3}$ and the binding energy at saturation density
$E_0= -16~$MeV. For the relativistic "cov" codes RVUU and DJBUU,
we employ a non-linear $\sigma-\omega$ Relativistic Mean Field (RMF) parameterization, with $M = 938.0~$MeV, 
$m_\omega = 783.0~$MeV, 
$m_\sigma = 550.0~$MeV, 
and the parameters $g_\sigma$, $g_\omega$, $A$ and $B$ given in Table \ref{tab:param}.
Since there are four free parameters, 
in addition to saturation density ($\rho_0$), energy per nucleon and compressibility at $\rho_0$, one can also fix the value of the Dirac mass $m^*$ at 
$\rho_0$. 
The parameterizations listed in Table II lead to the same values for  
compressibility, saturation density and binding energy 
as above, but with different values of the Dirac mass $m^*$.
\begin{table}[htbp]
\caption{\label{tab:param} Examples of RMF parameterization sets that give the 
required nuclear matter properties (see text).}
    \begin{tabular}{cccccc}
    \hline
    \hline
    Set & $m^*/M$ & $g_\sigma$ & $g_\omega$ & $A (fm^{-1})$ & $B$ \\
    \hline
  1 &  0.6 & 10.047638 & 12.247145 & -1.147188  & 12.396194\\
2 & 0.7 & 8.652969 & 10.346869 & -10.825788 & 75.221535\\
3 & 0.8 & 6.645764 & 7.953129 & -43.850479 & 277.549711\\
4 & 0.85 & 4.884545 & 6.411573 & -85.063619 & 461.632842\\
5 & 0.9 & 1.609514 & 4.340011 & -74.404620 & 172.583219\\
    \hline
    \hline
    \end{tabular}
\end{table}

In the SMASH code, two contributions to the vector field are considered: 
an attractive linear
field ($\beta_1$ = 2) with $C_1 =  -105.716/\rho_0$ MeV fm$^3$
and a stiffer repulsive field with $\beta_2 = 3.587$ and 
$C_2 = 52.836/\rho_0^{\beta_2-1}$ MeV fm$^{3(\beta_2-1)}$.
Finally, in pBUU the following parametrization is employed for the potential
$U(\rho_S)$: $a = -104.444$ MeV, $b = 43.0838$ MeV and $\sigma$ = 3.07326. 
In both  SMASH and pBUU, 
the model parameters have been selected to give  
the same compressibility, saturation density and binding energy as indicated above. 

\begin{figure}
\includegraphics[scale=0.55,angle=0]{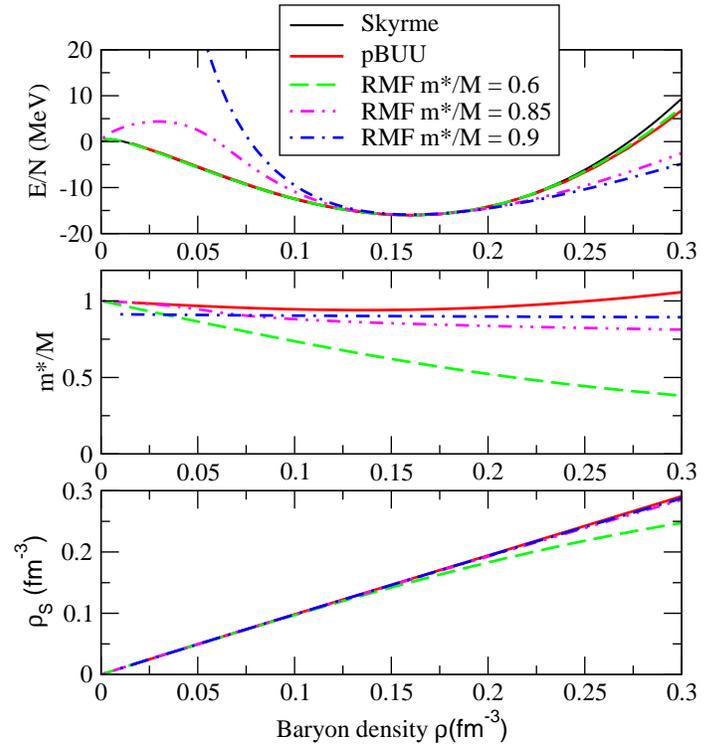}
\caption{Top panel: Energy per nucleon, as obtained for the adopted Skyrme-like
parametrization (full (black) line), three RMF parametrizations adopted 
for RVUU and DJBUU models and for the pBUU model (thick full (red) line). 
The lines for the Skyrme, pBUU and RMF $m^*/M = 0.6$ models strongly overlap.   
Middle panel: The Dirac mass $m^*$ as a function of the baryon density for pBUU and the three RMF  parametrizations. Bottom panel: The scalar density $\rho_S$ as a function of the nucleon density. }
\label{eos_analytical}
\end{figure}

In Fig.\ref{eos_analytical} (top panel), we show the energy per nucleon, $E/N$, 
for nuclear matter at zero temperature, as given by the 
adopted Skyrme parametrization (non-relativistic kinematics is considered, but
very similar results are obtained in the relativistic case and for SMASH), 
the parametrization
employed in pBUU and three RMF parametrizations, namely set 1, 4 and 5 of Table II.  One can see that all the curves shown in the panel exhibit the same trend
around saturation density, as expected.  Moreover, the pBUU curve is very close
to the Skyrme one in the whole density range considered. This is also the case for
the RMF parametrization with $m^*/M (\rho = \rho_0) = 0.6$.
For larger $m^*$ values, the curves deviate increasingly more from the Skyrme 
parametrization away from saturation density. However, for density variations
of about 20$\%$, as considered here, the differences are not large. 

The Dirac mass, $m^*/M$, is shown as a function of the density in the middle panel, whereas the bottom panel shows the density dependence of the scalar density,
$\rho_S$. Results are shown for all parametrizations considered in the top
panel, except the Skyrme interaction, which has no scalar field. 
One can see that the Dirac mass remains quite close to the nucleon mass, $M$,
over the whole density region considered, in the case of pBUU and of the
RMF parametrization with  $m^*/M (\rho = \rho_0) = 0.9$.

\begin{figure}
\vskip 1.cm
\includegraphics[scale=0.35,angle=0]{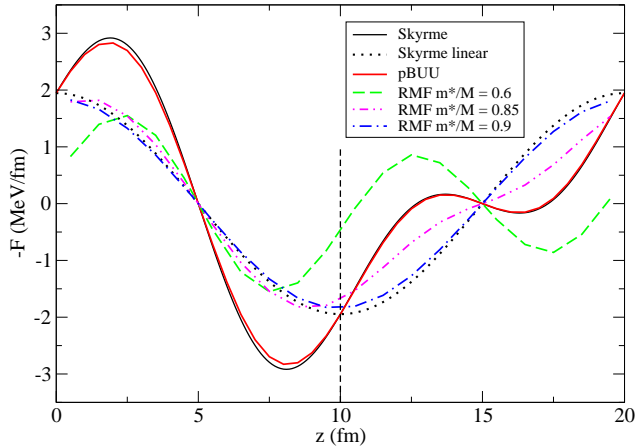}
\caption{The gradient of the mean-field potential, corresponding to
 the initial density standing wave (see text), as a function
of the $z$ position, as obtained for the adopted Skyrme-like
parametrization (full (black) line), three parametrizations for RVUU and DJBUU models and for the pBUU model (full (red) line). The dotted line corresponds to the
Skyrme interaction taken at frozen density ($\rho_0 = 0.16~fm^{-3}$), see text
for details. The short-dashed vertical line indicates the central position
of the box, where the density is equal to $\rho_0$ and the two Skyrme curves
cross.}
\label{gradient_analytical}
\end{figure}

Figure \ref{gradient_analytical} shows the corresponding gradients of the mean-field
potential, namely the quantity $-F(z) = \partial\epsilon / \partial z$, 
with $\epsilon$ being the single-particle energy,
calculated analytically for the initial standing wave impressed on the density profile. 
We note that according to the general definition of the single-particle energy, the quantity $F$ also depends on the momentum (for models including a scalar
field), thus we take the average of $F$ over the initial Fermi-Dirac momentum
distribution in this case. Namely, for the RMF parametrizations, we consider the quantity: 
\begin{equation}
-F(z) = \Bigl[\Bigl\langle\frac{m^*}{\sqrt{\vec{p}^2 + {m^*}^2}}\Bigr\rangle
\frac{dm^*}{d\rho_S}\frac{d\rho_S}{d\rho} + \frac{dA^0}{d\rho}\Bigr]\frac{d\rho}{dz}, 
\label{force_general}
\end{equation}  
 where the average is over momentum space and $\frac{d\rho}{dz}$ is the derivative of the initial sinusoidal perturbation. 
The same expression holds for pBUU, but with $A^0 = 0$.
In the case of the Skyrme interaction, namely for codes of type ``rel'' 
and ``non-rel'', and
for SMASH, one can simply write $-F(z) = \frac{dU}{d\rho} \cdot \frac{d\rho}{dz}$, where $U(\rho)$ is the corresponding mean-field potential. 
As one can see in the figure, though all parametrizations give the same
trend for the EoS around saturation density, quite interesting differences
exist for the gradient of the mean-field potential. This simply stems 
from the fact that different effective interactions may lead to the same EoS. 

Let us comment first the behavior associated with the Skyrme interaction, 
which is simpler to interpret. Within the linear regime, i.e., for very small
amplitude density perturbations, one can write  
$-F(z) = \frac{dU}{d\rho}|_{\rho=\rho_0} \cdot \frac{d\rho}{dz}$, and a
cosinusoidal trend would be obtained (see the dotted line in the figure).    
Thus, the behavior observed (full (black) line) can be ascribed to the amplitude
of the initial perturbation considered, which is not small and will induce non-linear effects in the
Vlasov dynamics. As it will be discussed in the following, mode coupling effects
are expected to appear. The behavior of the pBUU curve is very similar to the
Skyrme one. 
Turning to the behavior of the RMF parametrizations, we observe significant
differences with respect to the Skyrme interaction. It is interesting to 
notice that the parametrizations with large values of $m^*/M$ 
exhibit a trend close to the cosinusoidal one, indicating that the 
non-linearities introduced by the scalar field parametrization do not have
large effects on the gradients. 
It follows that, within the linear
regime,  these parametrizations (especially the one with $m^*/M (\rho = \rho_0) 
= 0.9$) are close to the behavior of the Skyrme interaction. The same does not hold
for the parametrization with $m^*/M = 0.6$. 
 We will show that  in spite of the presence of non-linear 
effects, the oscillation frequency of the initial density perturbation is
mainly determined by the features connected to the linear regime and the pure zero-sound propagation; thus we expect close results between the covariant codes
employing $m^*/M(\rho = \rho_0) = 0.9$ and the other codes. 
This point will be better illustrated in the following section.

We note that in the first formulation of the present homework,  a mean-field parametrization corresponding to the more realistic compressibility $K_0$ = 240 MeV was employed, as in the earlier comparisons of Au+Au collisions 
in Ref.\cite{Xu2016}.  The quite large damping effects observed in this case,
especially in QMD codes, made the analysis of the results not very transparent.  In order to get more persistent density oscillations, the homework was reformulated with the use of a nuclear potential corresponding to the larger, 
although unphysical,  
compressibility value, $K_0$ = 500 MeV.
 
\subsubsection{Details of the simulations and output of the codes} 
We have considered 10 runs for BUU-like codes, employing 
100 TPs per nucleon and 200 runs for QMD-type codes.
However, we should mention that to improve the quality of energy conservation
and momentum distribution features, the TP number was increased
for the codes that use point-like TPs, or 
triangles with $l$ = 1 fm, namely BUU-VM, IBUU and RVUU (see Table I).  
In particular, $N_{TP}$ = 1000 was adopted for IBUU {and pBUU}, 
and $N_{TP}$ = 2000 for BUU-VM and RVUU. 
For a reduced number of events, we output the (test) particle coordinates and
momenta at certain times in the evolution.
The main outputs of these calculations are tables of the average density 
$\langle\rho(z,t)\rangle$ and of the associated variance, reported as a function of the $z$ coordinate. 

More precisely, a grid along the $z$ direction, of size $\Delta z$, is introduced inside the box. We adopt $\Delta z$ = 1 fm.
For each event, the density $\rho(z,t)$, averaged over the $(x,y)$ plane,
is evaluated on the grid at each time step. Then the density is further 
averaged over all events and the associated variance is also 
evaluated. In the following, we omit the notation of the average.
For each event, we also calculate the gradient $-F(z)$
of the mean-field potential
along the $z$ direction, but only at the initial time t = 0. 

We will see in the following that the evaluation of the gradient of the mean-field potential is very helpful to understand the possible sources of discrepancies for the propagation of the density oscillations among the different codes.  
We also emphasize that from such a detailed output, it is possible to perform a Fourier analysis of the density oscillations, in space and in time, with sufficient accuracy.

The participating codes in this homework were given in Table I. The GiBUU code only contributed to the calculations with $K_0$ = 240 MeV (not shown here), 
and the DJBUU code only with $K_0$ = 500 MeV.

\subsubsection{Fourier transforms}
\label{sec:hw_Fourier}

To characterize the density perturbation introduced in the
initial conditions and its time evolution, it is useful to perform a Fourier analysis of the
density oscillations. We define the Fourier transform of the averaged spatial density as
\begin{equation}
 \rho_k(t) = \int_0^{L_z} {dz\, \rho}(z,t)\, \sin(kz),
\label{strength}
\end{equation} 
which gives a more compact representation of the spatial density
oscillations and can be called the strength function of the mode $k$. 
One generally observes damped oscillations as a function of time 
for the latter quantity. Ideally at the initial time, $t$ = 0, only the $k$ value 
corresponding to the initial perturbation, $k_{ini} = n\,2\pi/L_\alpha$ (with $n=1$
), leads to non-zero $\rho_k(t=0)$.  However, due to  fluctuations in the initial configuration, small admixtures of other modes can already appear at $t=0$. 
As time evolves, other $k$ components appear significantly. This can be called mode-mixing, which is due to the non-linear character of the Vlasov equation, but also to fluctuations.
For this reason, it is interesting to introduce also the Fourier transform of
the type: 
\begin{equation}
 \rho'_k(t) = \int_0^{L_z} {dz\, \rho}(z,t)\, \cos(kz),
\end{equation} 
and finally the quantity: $\rho_{k,tot}(t) = \sqrt{\rho_k^2(t) +  {\rho'_k}^2(t)}$. 

A deeper insight into the frequency and the damping of the density oscillations
is obtained from a further Fourier analysis of $\rho_k(t)$ with respect to 
time, i.e., the response function.  
Hence we introduce the quantity
\begin{equation}
{\rho}_k(\omega) =  \int_{t_{in}}^{t_{fin}} dt \, \rho_k(t)\, \cos(\omega (t-t_{in})),
\end{equation}
where the integration is extended over a time interval $\Delta t_{fi} = t_{fin} - t_{in}$, with a suitable choice of the initial time $t_{in}$ (see Sect. 
\ref{sec:response}). 
It is convenient to parametrize the frequency $\omega$ as 
$\omega = n_\omega \, \pi / \Delta t_{fi}$, where $n_\omega$ is an integer. 

\section{Analytical expectations for zero-sound propagation } 
\label{sec:Analytic}
In the idealized situation of a box calculation, it is possible to 
make analytical predictions for the
density oscillation frequency in the small-amplitude limit, according to the Landau theory of Fermi 
Liquids as applied to the linearized Vlasov equation \cite{linear}. 
Within the general relativistic framework introduced above, 
at zero temperature and density $\rho$, the zero-sound dispersion relation, which allows one to 
determine the oscillation frequency $\omega_k$ for the wave number $k$, 
reads \cite{Greco03,Matsui}:
\begin{equation}
 1 + {\tilde F_0} \phi(s) = 0,
\label{dispers} 
\end{equation}
where an effective Landau parameter, ${\tilde F_0}$, has been introduced 
and  $\phi(s)$ is the
Lindhard function: $\phi(s) = 1 - (s/2)~ \ln[(s+1)/(s-1)]$. 
The quantity  $s = \omega_k / 
(k v^*_F)$ represents the
sound velocity ($v_s(k) = \omega_k/k$) in terms of the Fermi velocity 
$v^*_F = p_F/E^*_F$.  
The energy 
$E^*_F = \sqrt{p_F^2 + {m^*}^2},$ where $p_F$ represents the Fermi
momentum, coincides with the Landau effective mass. 
Extending the results derived in \cite{Greco03,Matsui} to the more general case
of non-linear scalar and vector fields, 
the Landau parameter takes the following expression:
\begin{equation}
{\tilde F_0} = {{E^*_F} \over {3p^2_F}} [K^{pot} - 9 A_j \rho v^2_s].
 \label{Landau_rel}
\end{equation}
Here, $K^{pot} =  9\rho[f_\omega - f_\sigma{{m^*}^2 \over  {E^*_F}^2}(1+f_\sigma 
{\tilde A})^{-1}],$
with ${\tilde A} = 3(\rho_S/m^* - \rho/E^*_F)$, is the potential part of the
nuclear matter compressibility, and we have defined
$f_\omega = \frac{dA^0(\rho)}{d\rho}$ and $f_\sigma = \frac{d\Phi}{d\rho_S}$.
The term $\left( 9 A_j \rho v^2_s \right)$ inside the square bracket in Eq.(\ref{Landau_rel}) originates from the
spatial components of the nucleon four-current, and it is written in terms of
$A_j =  A^0(\rho)/\rho$
for RVUU and SMASH, and  $A_j = 0$ for all the other models.    
Making the approximation
$ v_s \approx v^*_F$, the frequency $\omega_k$ only appears inside the
Lindhard function, thus simplifying the solution of the dispersion relation. 
 
Eq.(\ref{dispers}) can be solved for all the models considered
here.  In particular, we note that for the models of the type ``non-rel'', i.e.,
in the non-relativistic limit ($v^*_F  = 
p_F/E^*_F \rightarrow p_F/M)$, the Landau parameter is written as
${\tilde F_0} = F_0 = \frac{3\rho}{2\epsilon_F}\frac{dU(\rho)}{d\rho}$, where 
the Fermi energy $\epsilon_F = p^2_F/(2M)$ has been introduced.  

Corresponding parameters and solutions for the sound velocity 
are reported in Table \ref{tab:disp}
for the different models. 
In the case of RVUU and DJBUU, 
several possibilities for the Dirac mass $m^*$ are included in the table.   
\begin{table}[htbp]
\caption{\label{tab:disp}
The Dirac mass $m^*$ (normalized to the nucleon mass $M$), the Landau parameter
${\tilde F_0}$, the solution of the dispersion relation, $s$, the quantity 
$(Mv^*_F)/p_F = M/E^*_F$ (see text), and the sound velocity, $v_s = \omega_k/k$, for the models
considered in the present work. 
}
    \begin{tabular}{cccccc}
    \hline
    \hline
    Type & $m^*/M$ & ${\tilde F_0}$ & $s$ & $M/E^*_F$ & $v_s$ \\
    \hline
  ``non-rel'' &  1 & 1.259 & 1.073 & 1 & 0.301\\
    \hline
   ``rel'' &  1 & 1.308 & 1.079  & 0.963 &  0.291 \\
    \hline
   ``cov'' & & & & & \\
   SMASH &  1   & 1.471 & 1.099 & 0.963 & 0.297 \\
   pBUU  & 0.942 & 1.208 & 1.067 & 1.017 & 0.304\\
   RVUU & 0.6 & -0.956 & - & 1.510 & -\\
   DJBUU & 0.6 & 0.496 & 1.005 & 1.510 & 0.425\\
   RVUU & 0.7 & -0.207 & - & 1.326 & -\\
   DJBUU & 0.7 & 0.704 & 1.017 & 1.326 & 0.378\\
   RVUU & 0.8 & 0.437 & 1.003 & 1.180 & 0.332\\
   DJBUU & 0.8 & 0.915 & 1.036 & 1.180 & 0.343\\
   RVUU & 0.85 & 0.728 & 1.019 & 1.117 & 0.319\\
   DJBUU & 0.85 & 1.022 & 1.047 & 1.117 & 0.328\\
   RVUU & 0.9 & 1.002 & 1.044 & 1.061 & 0.311\\
   DJBUU & 0.9 & 1.130 & 1.058 & 1.061 & 0.315\\

    \hline
    \hline
    \end{tabular}
\end{table}

The results obtained for the sound velocity, $v_s = \omega_k/k$, 
are closely related to the value of the 
Landau parameter ${\tilde F_0}$
and also of the Landau effective mass $E^*_F = p_F / v^*_F$.
For instance, for the models of type ``non-rel'' and for the mode that we are considering
($n=1$, $k = 0.314~fm^{-1}$), we have $\hbar\omega_k$ = 18.65 MeV. 
 
Zero-sound solutions are found only for $F_0 > 0$. The robustness of the
solution, $s =  \omega_k/(kv^*_F)$, of the dispersion relation increases 
with $F_0$, i.e., for larger compressibility values, as expected. 
Moreover, for a given solution $s$, a larger sound velocity is
obtained for larger values of the Fermi velocity $v^*_F$, i.e., smaller Landau 
effective mass.  
From Table \ref{tab:disp} one can see that for the considered compressibility
value, $K_0 = 500~MeV$, in the case of RVUU,
zero-sound solutions are obtained only if the Dirac effective mass
exceeds a threshold value, which is in the range $m^*/M = 0.7-0.8$.
Moreover, the Landau parameter ${\tilde F_0}$ is always larger in DJBUU than in 
RVUU. This behavior originates from the second term inside the
square bracket of  Eq.(\ref{Landau_rel}), which vanishes
in the DJBUU case due to its neglect of the spatial current $\vec{j}$ and
is negative in the RVUU case ($A^0(\rho) = f_\omega \rho$ there).
However, the sound velocity is similar in the two models and approaches
the values associated with the other models, if one considers large Dirac mass
values, see in particular the results obtained for $m^*/M = 0.9$.
This reflects the findings, illustrated in Fig.\ref{gradient_analytical}, that for the choice $m^*/M = 0.9$, the gradient of the mean-field potential is close to the trend given by 
the Skyrme interaction within the linear regime.     
Thus, in the following we will mainly consider this parametrization
(set number 5 in Table II).   

In the case of SMASH, the second term inside the square bracket of  
Eq.(\ref{Landau_rel}) is positive (because $A^0(\rho)$ is negative), 
leading to the large $F_0$ value reported in the Table. 
However, since the Landau effective mass is larger than the nucleon mass
in this case ($E^*_F/M$ = 1.038), the sound velocity turns out close
to the one associated with the other models.

To summarize our findings about the sound velocity $v_s$, 
one can say that relative to the models
of type ``non-rel'', the largest negative deviation is given by the 
``rel'' models (about -3 $\%$), 
whereas the largest positive deviation corresponds to the DJBUU model    
(about 4 $\%$, taking the parametrization with $m^*/M = 0.9$).

\subsection{Structure of zero-sound modes}
\label{structure}
{
In the following, we give more details about the structure of the zero-sound modes, which
can be deduced from the linearized Vlasov equation. For the sake of
simplicity, we present the formalism corresponding to the models 
of type ``non-rel'' and ``rel'', for which the derivation is straightforward. 
After performing a Fourier transform in space and time, the linearized
Vlasov equation 
can be expressed as \cite{AyikZFF}: 
\begin{equation}
f_k(\vec{p},\omega) =  \frac{\partial f_{FD}(p,T)}{\partial p}  
\frac{dU}{d\rho}  \frac{\cos(\theta_p)\rho_k(\omega)}{v\cos(\theta_p)
- \omega/k},
\end{equation}
where $f_k(\vec{p},\omega)$ represents the perturbation of the
distribution function associated
with the wave number $k$ and the frequency $\omega$.
The angle $\theta_p$ refers to the angle between the wave propagation direction
(namely the $z$ axis) and 
the momentum $\vec{p}$, and $v = \partial E/\partial p$. 
The self-consistent condition 
\begin{equation}
4\int \frac{d^3p}{(2\pi)^3} f_k(\vec{p},\omega) = 
\rho_k(\omega)
\label{disp_rel}
\end{equation}
leads to the dispersion relation discussed in the previous section, from which the collective solutions, $\omega = \pm\omega_k$, are derived. 
The corresponding zero-sound amplitude for a standing-wave solution 
of the distribution function at the initial time can be written as
\begin{equation}
f_k(\vec{p},t=0) 
= \frac{1}{2}[f_k(\vec{p},\omega_k) + f_k(\vec{p},-\omega_k)].
\end{equation}

On the other hand, in the homework calculations (performed at zero temperature), a spherical local Fermi
surface is chosen for the initial condition of the phase-space distribution, which can be expressed as:
\begin{eqnarray}
f(z,{p},t=0) = ~~~~~~~~~~~~~~~~~~~~~~~~~~~~~~~~~~~~~~~~~~~~~~~~~~~~~\\\nonumber
\theta(p_F-p) + f^{sph}({p},t=0)\sin(kz) \equiv \theta\Bigl(p_k^{\text{sph}}(z)-p\Bigr).
\label{f:def}
\end{eqnarray}
Here the zero temperature Fermi-Dirac distribution has been introduced 
implicitly, $f_{FD}(p,T=0)\equiv\theta(p_F-p)$,
and the function $p_k^{\text{sph}}(z) = p_F[1+(\rho_k\sin(kz))/\rho_0]^{1/3} \approx
 p_F[1+(\rho_k\sin(kz))/(3\rho_0)]$ describes the local spherical Fermi surface. 
By Taylor expanding the r.h.s. of Eq.(25), we obtain: 
\begin{equation}
f^{sph}({p}) = -\frac{\partial f_{FD}(p,T=0)}{\partial p} \cdot 
\frac{p_F}{3\rho_0} \cdot \rho_k 
\end{equation}
In this case, the amplitude, $\tilde{\rho_k}$, of the resonant 
density oscillations, 
associated with the collective zero-sound mode, 
can be recovered by
projecting the perturbation $f^{sph}(p)$ onto the auxiliary function, 
\begin{equation}
Q_k(\vec{p},\omega) \equiv  \frac{\omega/k}{v\cos(\theta_p)-\omega/k}~,
\end{equation}
which is recognized as the usual RPA amplitude \cite{AyikZFF}. 
Hence, we get
\begin{equation}
\frac{\tilde{\rho_k}}{\rho_k} = \frac{\langle Q_k(\omega_k)|f^{sph}\rangle}
{\langle Q_k(\omega_k)|f_k(\omega_k)\rangle}
+ \frac{\langle Q_k(-\omega_k)|f^{sph}\rangle}{\langle Q_k(-\omega_k)|f_k(-\omega_k)\rangle},
\label{projection}
\end{equation}
where the inner product stands for an integration over ${\vec p}$. 
At zero temperature, the integrals appearing in Eq.(\ref{projection}) can be solved analytically. We obtain
\begin{equation}
 {\langle Q_k(\pm\omega_k)|f_k(\pm\omega_k)\rangle} = \rho_k\Bigl(\frac{F_0}{s^2-1} - 1\Bigr)
\end{equation}
and
\begin{equation}
\langle Q_k(\pm\omega_k)|f^{sph}\rangle = \rho_k(1/F_0 + 1).
\end{equation}
Exploiting the values of $F_0$ and $s$ listed in Table III, we find
\begin{equation}
\frac{\tilde{\rho_k}}{\rho_k} = 
2\cdot\frac{1/F_0 + 1}{{F_0}/(s^2-1) - 1} = 0.49
\end{equation}
for the models of type ``non-rel'' and $0.51$ for the models of type ``rel''.
This means that only about half of the initialized perturbation is actually
a pure $n=1$ zero-sound mode. 
}

\section{Exact solution of the Vlasov equation: Locally Deformed Fermi Surface}
\label{sec:DFS}
\begin{figure*}
\includegraphics[scale=0.4,angle=0]{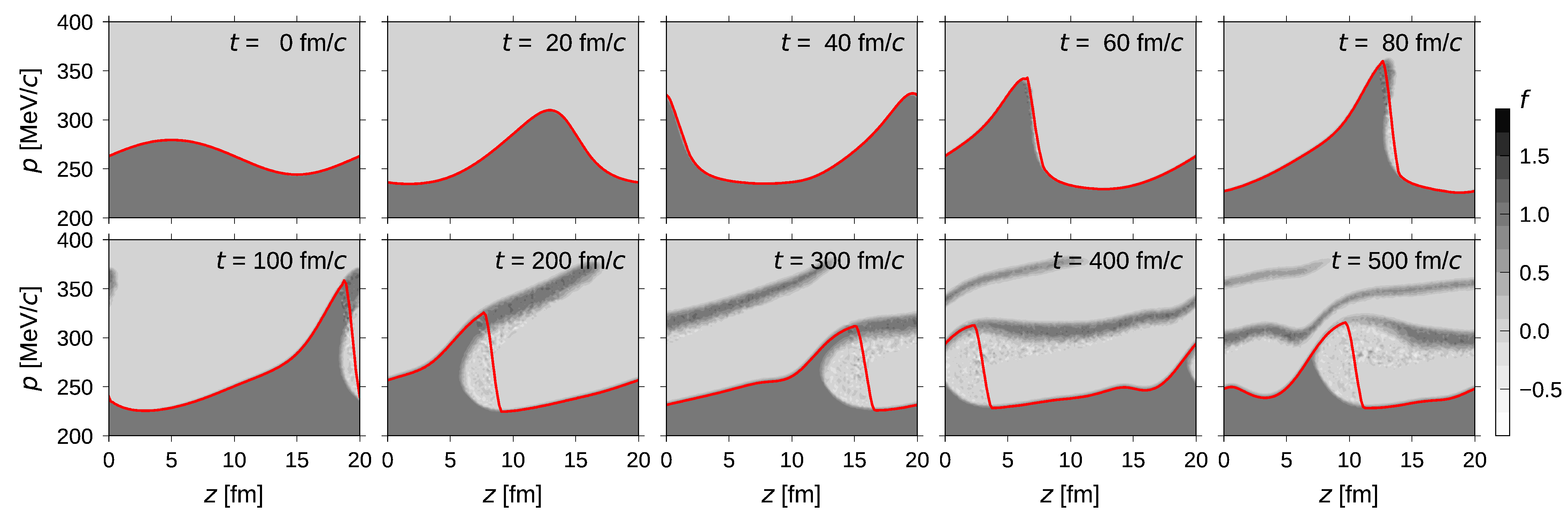}
\caption{
The phase space distribution $f$ in the plane of $z$ and
$p=|\vec{p}|$, averaged over the forward angle region $0 < \theta_p < \pi/16$, is
respresented at several time instants indicated in the different panels.  
The distribution $f$ is the sum of the two terms on the right hand side of Eq.
(\ref{weight}), and the function $p_{surf}(z,t)$ in the first term is represented
by the (red) solid line (for $\theta_p$ = 0).}
\label{DFS_results}
\end{figure*}

{
While we were able to derive exact limits for the oscillation frequency of the zero-sound mode in the small amplitude limit in the previous section, the further evolution of the wave is not analytic because of the non-linearity of the Vlasov equation, even when no fluctuations are present.
Hence, it is useful to have a direct (numerical) exact 
solution of the kinetic equation, for the general case of finite amplitude
and for the initial conditions of the homework, which are more general than
a pure zero-sound mode.  
These calculations are explained in this section and will be compared, in the
following, to the simulations of the transport codes.   
Owing to the axial symmetry of the 
simplified system that we are considering,
and the Liouville theorem for the given initial condition,
the nucleon distribution function can be represented in terms of axially symmetric deformations of the local Fermi sphere, which will be referred to as the
Deformed Fermi Surface (DFS) model in the following.    
From the Vlasov equation, a kinetic equation can be derived for the radius
of the deformed Fermi sphere as described below.  For the sake of simplicity, 
we will limit our considerations to the case where the single particle
energy is given as $\epsilon = \sqrt{\vec{p}^2 + M^2} + U(\rho)$
(as in the codes of type ``rel''), which also allows a straightforward extension
to the non-relativistic approximation in the ``non-rel'' case.  
 
Similar to the expression given by Eq.(25) 
for the initial distribution
function, the time-dependent phase-space distribution is written as 
\begin{equation}
f(z,p_z,p_\perp,t)=\theta\Bigl(p_{\text{surf}}(z,\theta_p,t)-p\Bigr),
\end{equation}
with $\tan\theta_p=p_\perp/p_z$ and $p=\sqrt{p_z^2+p_\perp^2}$. The function $p_{\text{surf}}(z,\theta_p,t)$, which describes the deformed Fermi surface, remains single-valued at least for a while from the initial time. From the Vlasov equation, the equation is obtained as
\begin{equation}
\frac{\partial p_{\text{surf}}}{\partial t}
+\frac{p_{\text{surf}}\cos\theta_p}{E(p_{\text{surf}})}
\frac{\partial p_{\text{surf}}}{\partial z}
-F(z,t)\biggl[
\frac{\sin\theta_p}{p_{\text{surf}}}\frac{\partial p_{\text{surf}}}{\partial\theta_p}+\cos\theta_p
\biggr]=0,
\label{eq:DFS}
\end{equation}
which can be easily solved numerically. 
In the above, $E(p_{\text{surf}})= {\sqrt{M^2+p_{\text{surf}}^2}}$ reduces to $M$ 
in the non-relativistic approximation.
The force is expressed as
\begin{equation}
F(z,t)
=-\frac{dU}{d\rho}\frac{\partial\rho(z,t)}{\partial z}.
\end{equation}
In the case of the homework condition, the Fermi surface $p_{\text{surf}}(z,\theta_p)$ becomes multi-valued after $t\approx 50$ fm/$c$. To handle such cases, test particles with positive and negative weights are introduced, so that the phase-space distribution is now written as
\begin{equation}
f(z,p_z,p_\perp,t)=\theta\Bigl(p_{\text{surf}}(z,\theta_p,t)-p\Bigr)+
\label{weight}
\end{equation}
$$
+\frac{1}{4}\sum_{k}W_k\delta(z-Z_k(t))\delta(p_z-P_{z,k}(t))
\frac{\delta(p_\perp-P_{\perp,k}(t))}{2\pi p_{\perp}}
$$
with the weights chosen to be
\begin{equation}
W_k=\pm\frac{1}{N_{\text{TP}}}\frac{(2\pi)^3}{L^2}, 
\end{equation}
and $N_{\text{TP}}=5000$. 
{Thus a test particle corresponds to a small fraction, $\pm\frac{1}{N_{\text{TP}}}$
of a full nucleon, which spreads uniformly on a plane perpendicular to the
$z$ axis and on an axially symmetric ring in the momentum space.}  
At every time step of $\Delta t=0.25$ fm/$c$, after considering
the evolution from Eq.(\ref{eq:DFS}) for the single-valued function $p_{\text{surf}}$ and the classical equation of motion for the existing test particles $(Z_k,P_{z,k},P_{\perp,k})$, the surface is replaced by its smoothed version, $p_{\text{surf}}\to\tilde{p}_{\text{surf}}$, and a suitable number of test particles are newly created to compensate for the change 
$\theta(\tilde{p}_{\text{surf}}-p) - \theta(p_{\text{surf}}-p)$. For each value of $\theta_p$, the smoothed version is defined by replacing the function in the region of $z\in[z_-,z_+]$ ($z_{\pm}=z_0\pm \frac{15}{32}\ \text{fm})$ around the point $z_0$ of the maximum of $|\partial p_{\text{surf}}/\partial z|$ with a polynomial fit using the three points at $z_-$, $z_0$ and $z_+$. In a similar way, the function is further smoothed for the variable $\theta_p$ for each $z$.
Results for the time evolution of the Fermi surface deformations, as obtained 
in the homework conditions, for the Skyrme interaction, 
are represented in 
Fig.\ref{DFS_results}.
The figure shows the phase space distribution $f(z,p_z,p_\perp,t)$ in the plane determined by
$z$ and $p=|\vec{p}|$, averaged for the forward angle region $0 < \theta_p < \pi/16$. 
One clearly observes that the Fermi surface 
is multivalued, corresponding to breaking waves, and  
eventually takes a ``millefeuille'' shape. 
}

\section{Density oscillations in a box: results from transport 
codes}
\label{sec:oscill}

After the standing wave has been initialized, it is propagated using the Vlasov part of the various transport codes. Here we will see significant differences, which to a large part tie to the fluctuations introduced by the chosen representation of phase space. As will be seen below, although the system is initialized as a Fermi system, its character changes in the evolution, 
with the degree of the change depending on the code family
and the individual codes. 
The consequences of these effects  will be studied in the following sections in terms of Fourier transform coefficients. 

\begin{figure}
\vskip 1cm
\includegraphics[scale=0.5,angle=0]{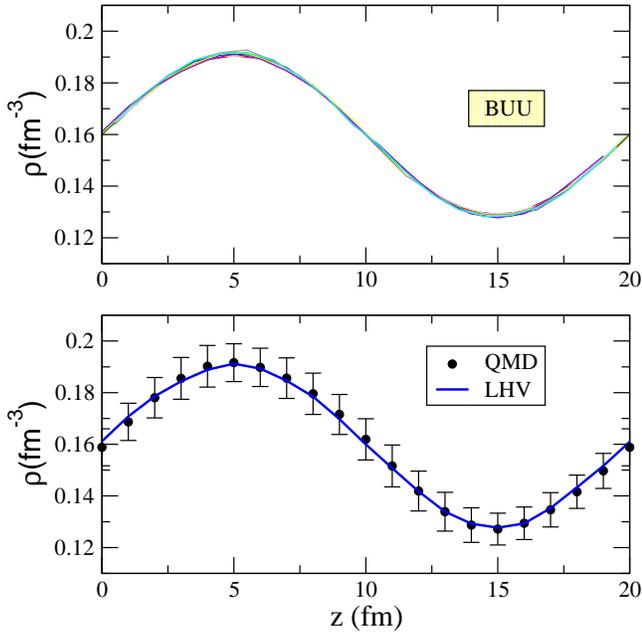}
\caption{Initial profiles of the standing wave as initialized by the BUU-type codes (top) and the QMD-type codes (bottom). The result corresponding to a
selected BUU-like code (LHV) is also shown on the bottom panel (full line) 
for comparison.}
\label{initializ_all_new}
\end{figure}

\begin{figure*}
\vskip 3.2cm
\includegraphics[scale=0.65,angle=0]{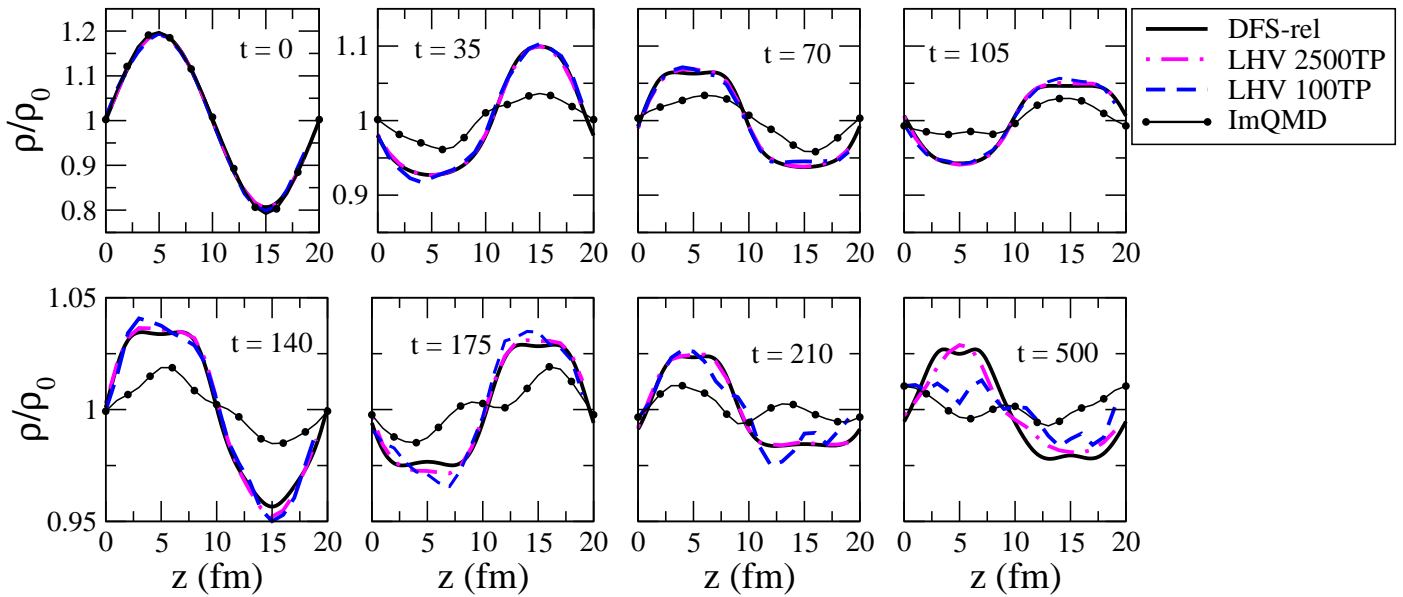}
\caption{Density for  the initialized standing wave at several instants of 
time, indicated in the panels in fm/$c$.
The results are shown for DFS calculations ((black) full line), LHV calculations
with 2500 TP per nucleon (dot-dashed (magenta) line) and 100 TP per nucleon 
((blue)
dashed line), 
and ImQMD calculations 
(full line with dots). }
\label{oscill_all}
\end{figure*}

\subsection{Coordinate space}
\label{sec:osc_coordinate_sp}

The initial average profiles in $z$-direction for all participating codes are shown in Fig.\ref{initializ_all_new}. Both BUU-like and QMD-like codes 
give a faithful initialization.  In the case of QMD-like codes,
the figure also shows the standard deviation of the density $\rho(z,t=0)$, as 
obtained from the sample of the 200 events considered.  
{The average agrees very well with the average trend associated with 
the BUU codes}. 
The standard deviation is
reduced by about a factor 10 for the BUU-like codes and is not shown on the figure. 
The evolution of the standing wave profile with time is shown in Fig.\ref{oscill_all}, for some representative codes and the DFS model. 
In particular, the results of DFS calculations (with relativistic kinematics) are compared
to the evolution of the density profile predicted by a selected BUU-like model (LHV) and a selected QMD-like
model (ImQMD).  In the case of LHV, in addition to the calculations corresponding to the homework
conditions (100 TP per nucleon), we also consider results obtained with a larger TP number, 
namely $N_{TP}$ = 2500.  

According to the features characterizing zero-sound propagation in nuclear
matter, we 
expect damping effects in the density oscillations, related to
the interplay between the collective response induced by the initial perturbation, the mode-coupling due to the non-linearity of the Vlasov equation,  and the disordered particle Fermi motion (Landau damping). 
One can appreciate 
the non-linearity of the system evolution from the distortion
of the original sinusoidal wave form. 
Moreover, in each of the simulated events, owing to the finite phase space mapping, numerical density fluctuations are present on top of the standing
wave initially introduced. These fluctuations act as an additional 
(numerical) source of damping. 
We recall that the smaller the number of TPs used, the larger the
amplitude of the 
density fluctuations. Indeed, for a given event, the density fluctuation variance 
can be expressed as 
\begin{equation}
\sigma^2_\rho = {\bar{\rho}_z}/(N_{TP}V), 
\label{var}
\end{equation}
with the volume $V$ 
typically associated with the extension of the nucleon (or TP) 
wave packet. 
In Eq.(\ref{var}), $\bar{\rho}_z$ represents the density averaged over the cells having position $z$.
The particles 
momentum distribution presents a similar kind of statistical fluctuation. 

A quite good agreement is observed between DFS and LHV calculations
with 2500 TP per nucleon (for which numerical fluctuations are negligible), up to the final time considered (t = 500 fm/c).  A reasonable agreement
is seen also with the calculation adopting 100 TP, though in this case a quenching of the 
oscillation amplitude appears at large times, clearly visible at $t_{fin}$ = 500 fm/c.  As expected, 
damping effects 
of the density oscillations are more pronounced in ImQMD calculations; in this case, the density profile starts to exhibit a random character already around t = 200 fm/c. 
As anticipated above, 
we conclude that the damping effects observed in LHV and in ImQMD, relative to DFS calculations,
are connected to the amount of fluctuations inherent to
the transport code family. 

 \begin{figure*}
\vskip 1.4cm
\includegraphics[scale=0.6]{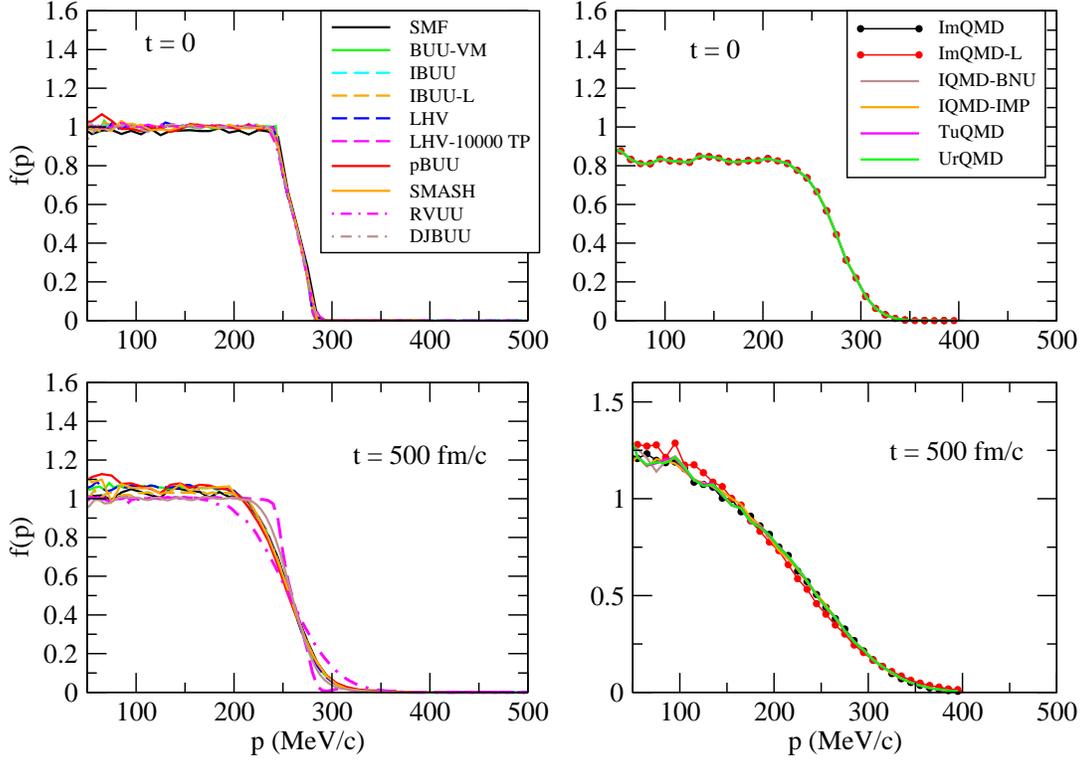}
\caption{
Momentum distributions in the different 
codes at initial (top panels) and final (bottom panels) times: BUU-like in the left panels and 
QMD-like in the right panels.
The QMD codes have used an identical initialization of the nucleon positions
and momenta.
}
\label{moment_all}
\end{figure*}

\subsection{Momentum distribution and energy conservation}
\label{sec:mom_energy}

{In Fig.\ref{moment_all} we show the distribution of the absolute value of the particle momentum, 
$f(p) = (2\pi)^3n(p)/(4V_{\text{ps}})$, where $n(p)$ is the number of nucleons 
with momentum $p$ and $V_{\text{ps}}$ represents the phase-space volume: 
$V_{\text{ps}} = V_{box} \cdot (4\pi p^2)\Delta p$. $V_{box} = L_\alpha^3$ denotes the 
volume of the box and we adopt $\Delta p$ = 5 MeV/c.
Results are shown in the left panels for BUU codes and in the right 
panels for QMD codes}. The distribution is shown for the initialized configuration in the top, and for the final configuration in the bottom panels. At the initial time, for homogenous matter at saturation density this would be a step function at the Fermi momentum of about 265 MeV/c. As observed for the BUU-like codes, there is a slight smearing, due to the impressed standing wave.  For the QMD-like codes, 
a considerably larger smearing is observed, corresponding to the larger intrinsic initial density fluctuations (generating a wider range of local Fermi momenta).  
It should be noticed that all QMD codes have employed exactly the same input for the
initialization. 

{ The results obtained by adopting the extreme choice of $N_{TP}$ = 10000 in LHV
calculations show that the initial momentum distribution should be approximately preserved in time. Indeed the final configuration is very close to the initial one. 
However, it is seen that in general the momentum distribution changes by amounts that depend on the code.} Most BUU codes reasonably well preserve the quantum-statistical behaviour.
The QMD-like codes in the right bottom panel are seen to deviate significantly from the Fermi statistics at the final time, approaching the classical Maxwell-Boltzmann distribution. 
This behavior can be ascribed to the larger fluctuations inherent to the QMD approach. 

\begin{figure}
\vskip 1.cm
\includegraphics[scale=0.35]{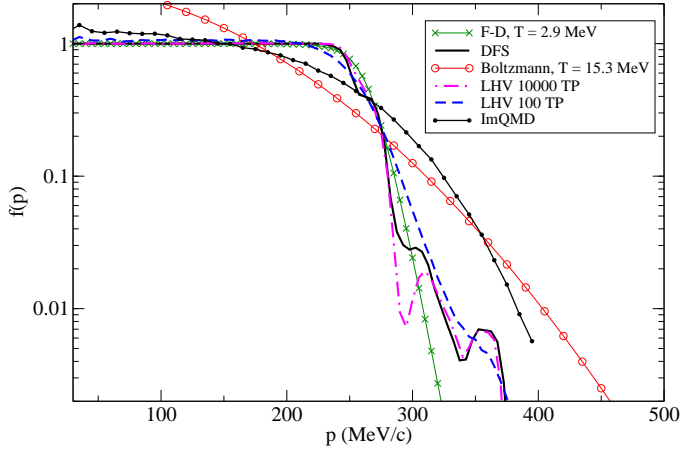}
\caption{
Momentum distributions, in log scale, at the final time t = 500 fm/c, as
obtained in LHV calculations with $N_{TP}=100$ (homework conditions, dashed
(blue) line) 
and $N_{TP}=10000$ (dot-dashed (magenta) line), DFS (full (black) line) 
and ImQMD calculations (full line with dots).
The lines with (green) crosses and (red) open dots represent the Fermi-Dirac
and Boltzmann distributions, respectively, at the temperature value corresponding to the system total energy. 
}
\label{fig_tail}
\end{figure}

{These features are better illustrated in Fig.\ref{fig_tail}, 
which shows the results of selected BUU- and QMD-like codes, 
compared to DFS calculations and to the trend associated with the
Fermi-Dirac and Boltzmann distributions
at the temperature value corresponding to the system total energy, in the fermionic
(T = 2.9 MeV) and the classical (T = 15.3 MeV) case, respectively.
As shown by the exact DFS calculations, and also by LHV calculations 
employing $N_{TP}=10000$, the Vlasov dynamics 
does not bring the system towards the finite 
temperature Fermi-Dirac distribution, as one would instead observe in 
the presence of two-body collisions. 
Features connected to the multi-valued structure
of the Fermi surface $p_{surf}(z,\theta_p,t)$ induced by non-linearities (see 
the discussion of DFS calculations in Sect.\ref{sec:DFS} and Fig.\ref{DFS_results})
appear
in the high momentum tail of $f(p)$.   
The signatures of the ``millefeuille'' structure 
shown in Fig.\ref{DFS_results} are clearly visible also in LHV calculations 
employing $N_{TP}=10000$, which indeed exhibit noticeable similarities with 
the DFS calculations.  
In LHV calculations with $N_{TP}=100$, the system moves slightly towards a classical behavior, as indicated by the fact that the
distribution function takes values slightly larger than $f(p)$ = 1 (see also Fig.\ref{moment_all}), with
the overall shape of
the momentum distribution reasonably well preserved.
The high momentum structures
are smeared out in this case. 
  
On the other hand, 
as already discussed above, significant deviations from the fermionic
behavior are observed for
the QMD codes, which tends to approach the Boltzmann distribution.


}

Finally, we mention that the total energy is conserved in all codes, 
within 
$1\%$ in the worst case.
The violation of energy conservation results from the numerical solution of 
Eq.(\ref{eq:prop}) in the coding process. Mostly, the Euler's method, 
the fourth order Runge Kutta method, and the leapfrog method are applied in the different transport codes. In principle, the numerical error is reduced 
when employing a higher-order method. To investigate the impact of 
the aforementioned numerical methods on the calculations considered
in this work, the fourth order Runge Kutta method (default method) and the Euler's method have been tested within the UrQMD model. 
It is found that UrQMD with the default method and UrQMD-Euler lead to convergent predictions, 
which are almost completely overlapping and thus not shown in the figures. 
However, it should be noticed that the excellent agreement between the two
methods is favored by the 
quite low excitation energy charactering the system considered. 


\section{
Illustrative results for 
selected codes}
\label{sec:selected}

As discussed in Sect.\ref{sec:hw_Fourier}, the damping and frequency effects can be more compactly seen in the Fourier transform coefficients with respect to 
coordinate space, called the strength function, and with respect to time, called the response function. These depend not only on the dynamical features of the Vlasov equation but also on the implementation in the specific codes, as we already saw in Sect.\ref{sec:oscill} .
To illustrate these effects and their dependence on the type of transport code, we will in this chapter compare in detail results of a selected BUU code of type ``non-rel'', 
namely SMF, a selected BUU code of type ``rel'', namely LHV, and a selected QMD-type code, namely 
ImQMD. As a reference, DFS results will also be shown.
We will, in particular, study how the results depend on approximations and calculational parameters of the codes. In the following section, we will then make this comparison for all participating codes. 


\subsection{Strength function}
\label{sec:selected_FT}

The frequency of the oscillation and the damping of the amplitude can be compactly seen in the behaviour of the first Fourier transform coefficient, 
$\rho_k(t)$, i.e., of the mode strength (Eq.({\ref{strength})). 
\begin{figure}
 \vskip 1.cm
\includegraphics[scale=0.35]{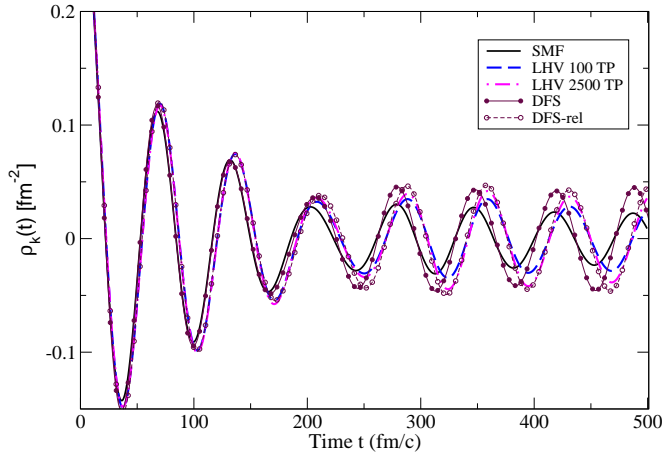}
\caption{
The Fourier transform coefficient, $\rho_k(t)$, for the node number $n$=1 is displayed as a function of time. 
LHV and SMF results are compared to DFS calculations with and without
relativistic kinematics. }
\label{DFS_strength_comp}
\end{figure}
This is shown in Fig.\ref{DFS_strength_comp}, where DFS calculations (with
and without relativistic kinematics) are compared to SMF and LHV calculations.  
In order to simulate the behavior of transport codes, in DFS the density $\rho(z,t)$, calculated by integrating $f(z,p_z,p_\perp,t)$ over the momentum, is smeared by a triangular distribution (extending to $\pm 2$ fm) and the derivative $\frac{\partial}{\partial z}$ is replaced by a finite difference (of the two points at $\pm 1$ fm).

We note that the value of the Fourier transform coefficient $\rho_k(t)$ at the initial time $t= 0$ is equal to $\rho_k(t=0) = (L_z/2) \, a_\rho $
= 0.32 fm$^{-2}$.
The early strong reduction of the oscillation amplitude seen in Fig.8 corresponds to the projection
of the momentum distribution of the initial perturbation on the zero-sound
mode, as discussed in Sect. \ref{structure}.
The subsequent behavior reflects 
damping and mode-coupling effects, as discussed below.

An excellent agreement with the density oscillation 
trend predicted by DFS, both for the non-relativistic and the relativistic version, is observed until t$\approx$250 fm/c. At later times
more pronounced damping effects, of numerical origin, are seen in the simulations. However, it is interesting
to notice that, when 2500 TP are employed in LHV, the simulations are very
close to the DFS-rel results up to the final considered time of t = 500 fm/c.

Now we move to discuss in more detail the impact of the main numerical 
ingredients of BUU- and QMD-like codes on the Vlasov dynamics.  
In Fig.\ref{FT_SMF-ImQMD}, we show in the upper panel results obtained from SMF calculations with different TP numbers and in the lower one from ImQMD calculations with different wave packet widths. 

 \begin{figure}[h]
\includegraphics[scale=0.35]{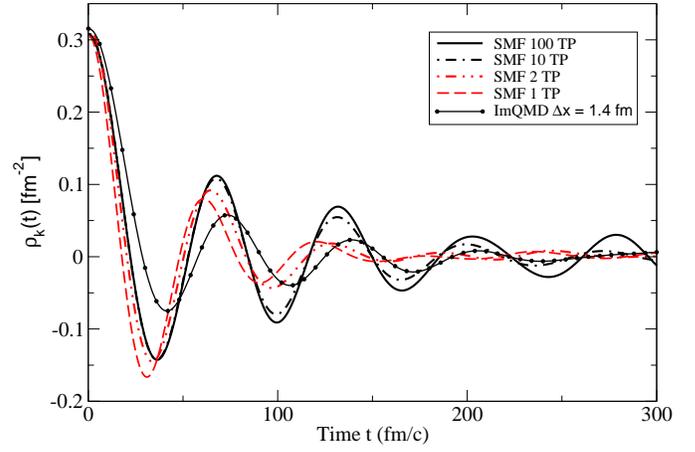}
\vskip1.3cm
\includegraphics[scale=0.35]{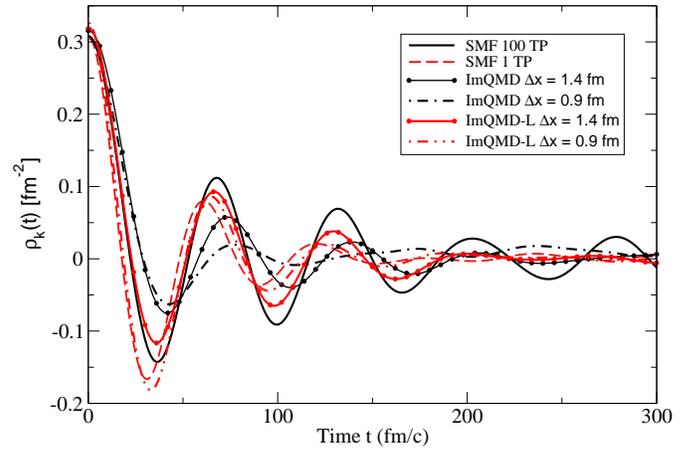}
\caption{
The first Fourier transform coefficient, $\rho_k(t)$, as obtained for an initial sinusoidal perturbation with node number $n$=1, as a function of time. In the 
top panel, SMF calculations are shown employing different numbers of TPs, as indicated in the legend. Moreover, standard ImQMD calculations are represented. 
In the bottom panel, ImQMD and ImQMD-L calculations are shown employing different Gaussian wave packet widths, and for comparison the SMF calculations with 100 and 1 TP per nucleon.}
\label{FT_SMF-ImQMD}
\end{figure}

When employing 100 TPs in the SMF calculation, the statistical fluctuations according to Eq.(\ref{var}) are quite suppressed, and the numerical damping remains limited.  On the other hand, one can nicely observe that owing to fluctuations, the damping increases strongly
 when using a smaller number $N_{TP}$ 
in the calculations (see the behavior for $N_{TP}$ = 10 and  $N_{TP}$ = 2). 
At the same time, the oscillation frequency is seen to slightly
increase. 
It is also interesting to observe that SMF results with $N_{TP}$ = 1 are different from molecular dynamics calculations, as reported in the figure for 
the ImQMD results.  In the following, we will explore the reasons
for this behavior in more detail. 

The influence of fluctuations in the context of QMD-like models is
investigated by considering
ImQMD calculations that employ, in addition to the standard value of the
Gaussian width ($\Delta x = \sqrt{2}$ fm), another choice, namely 
 $\Delta x$ = 0.9 fm.  This Gaussian width has been selected to fit
the width of the triangular profile employed in SMF (see Table I). 
Moreover, also the Lattice version of ImQMD  (ImQMD-L) is considered.
The corresponding results are shown in the lower panel of
Fig.\ref{FT_SMF-ImQMD}.

Comparing the (black) dot-dashed and full (with circles) lines, 
one can see that
reducing the Gaussian width in ImQMD (approaching the width value employed 
in SMF) 
leads to quite quenched oscillations, thus increasing the discrepancy with the
corresponding 
SMF results with 1 TP, contrary to what might have been expected.
On the other hand,   
quite interesting results are seen for ImQMD-L: 
in this case, calculations with the reduced width, $\Delta x$ = 0.9 fm, are
quite close to SMF results with  $N_{TP}$ = 1. Moreover,    
employing the standard value of  $\Delta x$ = 1.4 fm, the strength function
exhibits an oscillation frequency similar to standard SMF calculations (i.e., with $N_{TP}$ = 100), though with more pronounced damping effects. 
The results presented so far demand clarifications concerning the 
relation between the oscillation
frequency and the wave packet width in QMD-like approaches, or the test
particle number in BUU-like approaches, which will be given in the next subsection.  

 \begin{figure*}
\vskip3.2cm
\includegraphics[scale=0.7]{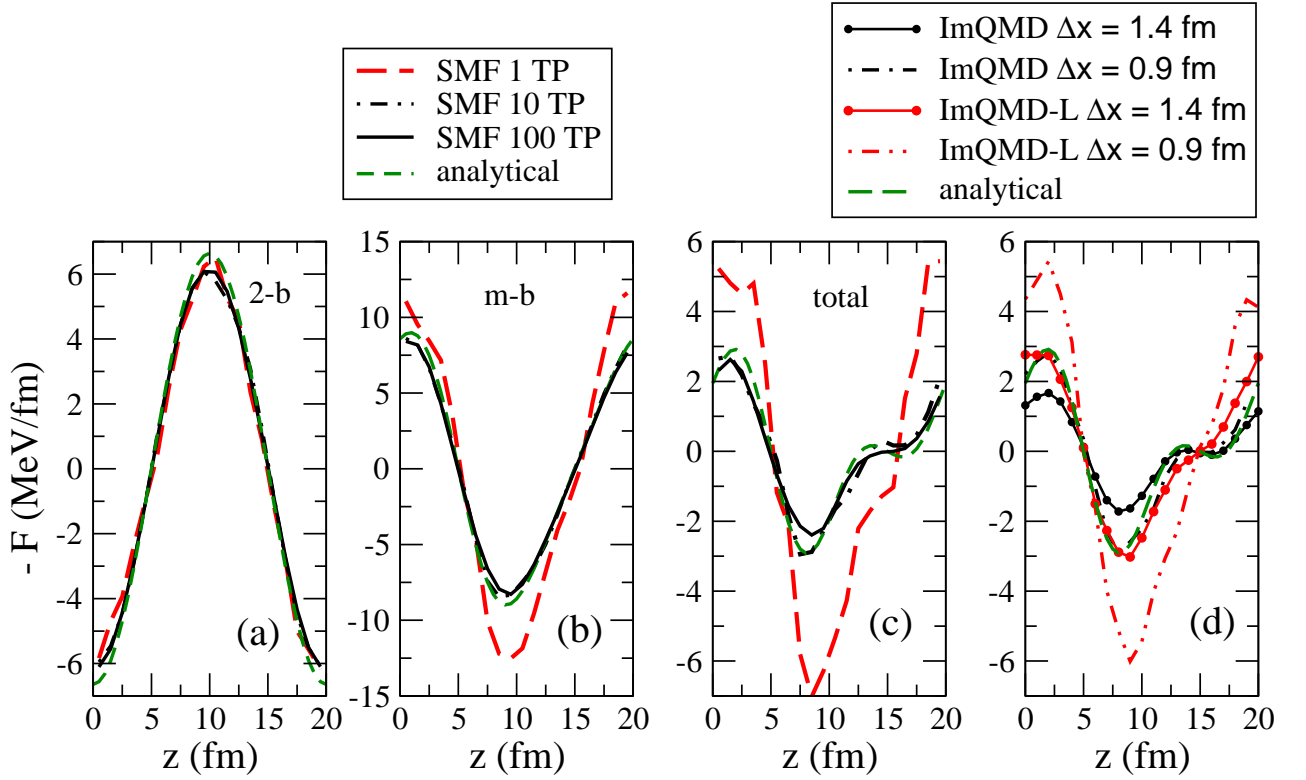}
\caption{
Gradient of the mean-field potential. Panels (a)-(c) correspond to SMF calculations, with several options for the TP number, from two-body (2-b), many-body (m-b) and total contributions, respectively, with the legend given by the left panel on top of the figure.  Panel (d) corresponds to ImQMD and ImQMD-L calculations using several options for the Gaussian width, with the legend given by the right panel on top of the figure. The analytical curve corresponds to 
the dashed (green) line. }

\label{gradz_SMF-ImQMD}
\end{figure*}

\subsection{Gradients of mean-field potential}
\label{sec:selected_gradZ}

The oscillation frequency crucially depends on the gradients of the mean-field
potential that the particles feel, and on the interplay with fluctuations. 
Indeed, the gradients determine the change of the momenta
of nucleons (or TPs) according to the equation (for codes of type ``non-rel'' 
and ``rel'')
\begin{equation}
\frac{\partial}{\partial t}{P_{z,i}} = -\partial \epsilon / \partial Z_i = - \partial U / \partial Z_i,
\end{equation}
where $\epsilon$ is the single-particle energy. 
As already discussed in Sect. \ref{nucl_int}, 
the gradient can
be calculated analytically at the initial time according to the perturbation
impressed on the system:
\begin{equation}
(\partial U / \partial z)_{t=0}
 =  1/\rho~ [a(\rho/\rho_0) + b\sigma (\rho/\rho_0)^\sigma] (a_\rho k \cos(kz)).
\end{equation} 
In the simulations, 
we have evaluated, for each event, the gradient,  $\partial U / \partial z$, 
of the mean-field potential
along the $z$ direction at the initial time t = 0. The codes calculate this quantity for each TP or nucleon.  Then, for each cell of the $z$ grid (with side
equal to 1 fm), 
this quantity is averaged over all nucleons (or TPs) having the $z$ coordinate inside the grid (i.e., within 1 fm of interval) for any value of the $(x,y)$ coordinates. 
Finally, we average over all events considered.
A plot of these gradients is shown in Fig.\ref{gradz_SMF-ImQMD}: 
in panels (a-c) for SMF for the two- and many-body parts and the total 
gradients with different TP numbers, respectively, and in panel (d) for ImQMD for the two
versions of the code and for different wave packet widths.  
One clearly observes that the gradients depend on the number 
of TPs employed (for SMF) and on the Gaussian width (for
ImQMD). In particular, as shown by the first two panels of Fig.\ref{gradz_SMF-ImQMD}, the gradient associated with the linear ($a\rho$) term of the mean-field
potential is not influenced by the TP number adopted, whereas a dependence
on the number of TPs is seen for the stiffer (many-body) $b\rho^\sigma$ term. 
This can be understood as follows: the gradient $ \partial U / \partial Z_i$
can be written as 
\begin{equation}
\frac{\partial U}{\partial Z_i} 
\approx \int d^3r ~ U(\rho)~ \frac{\partial G(\vec{r}-\vec{R}_i)}{\partial Z_i} = 
\frac{\partial H_{pot}}{\partial Z_i},
\label{ham_pot}
\end{equation}
where 
\begin{equation}
 H_{pot} = \int d^3{r}~ [\frac{a}{2}~(\rho^2/\rho_0) +
\frac{b}{\sigma+1}~(\rho^{\sigma+1}/\rho_0^\sigma)].
\label{H_pot}
\end{equation}
We consider the average of the middle part of Eq.(\ref{ham_pot}), by 
summing over the different cells
with the same position on the z axis. 
Starting from the definition of the mean-field potential, $U(\rho)$, 
it is easy to realize that one has to deal with the 
average value of $\rho$ and $\rho^{\sigma}$.  
Thus, the linear ($a\rho$) term of the potential is not affected
by the fluctuations, whereas for the second ($b\rho^\sigma$) term one can write $\langle\rho^\sigma\rangle = {\bar \rho_z}^\sigma 
+ \frac{\sigma(\sigma-1)}{2}~{\bar \rho_z}^{\sigma-2}\sigma^2_\rho$.
Exploiting the expression of the variance, Eq.(\ref{var}), this 
quantity can be rewritten as
$\langle \rho^\sigma\rangle = {\bar \rho_z}^\sigma [1 + \sigma(\sigma-1)/(2{\bar \rho_z} V N_{TP})]$.
Thus the average gradient of the many-body term is affected by the presence of 
fluctuations, which affect the repulsive part of the nuclear 
effective interaction. 
In particular, the presence of fluctuations induces larger
gradients (in absolute value) with respect to the analytical predictions. 
This effect clearly appears in SMF calculations when decreasing the number of TPs,
as shown in Fig.\ref{gradz_SMF-ImQMD}(b).  

In particular, when using $N_{TP} = 100$ or even  $N_{TP} = 10$, 
fluctuations are quite
reduced and the average gradient follows the analytical predictions. 
On the other hand, considering just one TP per nucleon, the 
gradient gets larger values.  Similarly, in the case of ImQMD (panel (d)), 
smaller values of the Gaussian width (i.e. larger fluctuations) lead to
larger density gradients.  Confronting SMF calculations with $N_{TP} = 1$
with ImQMD results of similar width ($\Delta x$ = 0.9 fm), one can see that the 
latter gives a smaller gradient (which is accidentally close to the analytical 
curve). 

This result can be connected to an approximation, often
employed in QMD-like codes, to evaluate the gradients associated with the
many-body term of the Skyrme interaction.
Within QMD-like approaches and employing Gaussian functions for the nucleon
wave packet, the first term of the nucleon
potential energy can be written as
\begin{equation}
H_{pot}^{2body,QMD} = {a \over 2\rho_0}\sum_i {\tilde \rho}_i, 
\label{QMD_2b}
\end{equation} 
where ${\tilde \rho}_i$ is defined as
\begin{equation}
{\tilde \rho}_i = [4\pi (\Delta x)^2]^{-3/2} \sum_j exp[-(\vec{R}_i - \vec{R}_j)^2
/(4 (\Delta x)^2)]
\label{int_dens}
\end{equation}
Whereas the combination of Eq.(\ref{QMD_2b}) and Eq.(\ref{int_dens})  
yields the exact
two-body contribution to the Hamiltonian, a similar combination, 
\begin{equation}
H_{pot}^{3body,QMD} = {b \over (\sigma+1)\rho_0^\sigma}\sum_i {\tilde 
\rho}_i^\sigma,
\label{QMD_3b}
\end{equation} 
\noindent
does not yield the exact result for 
the stiffer repulsive term of the potential energy.

The approximation Eq.(\ref{QMD_3b}) leads to a reduction of the strength of the latter term and
seems to be the origin of the results observed in 
Fig.\ref{gradz_SMF-ImQMD}(d) for ImQMD. It will be seen below that this is also the case for the other QMD-like codes involved in the comparison.
However, it should be noticed that 
the Lattice formulation of ImQMD (ImQMD-L) is free from this problem, 
thanks to the exact calculation of the many-body term \cite{Limqmd}.    

This explains the results
shown in Fig.\ref{gradz_SMF-ImQMD}(d)
that ``standard'' ImQMD calculations, with $\Delta x$ = 
$\sqrt{2}\approx$ 1.4 fm, give a smaller gradient than the analytical predictions. This is
due to the approximation discussed above for the gradient and to the Gaussian width amplitude, 
which introduces smearing effects.  Reducing the width to $\Delta x$ = 
0.9 fm, the gradient increases (dot-dashed curve on the figure) and becomes
(accidentally) closer to the analytical value.  
In the case of ImQMD-L, owing to the different treatment of the stiff term
of the nuclear potential, the choice of $\Delta x$ = 1.4 fm yields
results that coincide with the analytical curve. On the other hand, with $\Delta x$ = 
0.9 fm, the trend approaches SMF results with 1 TP, as expected. 
These findings explain why, as seen in Fig.\ref{FT_SMF-ImQMD}, ImQMD-L calculations with  $\Delta x$ = 0.9 fm give density oscillations pretty close to the SMF
results with 1 TP, whereas ImQMD-L calculations with  $\Delta x$ = 1.4 fm 
yield results close to the SMF calculations with 100 TP. Indeed, the gradient of the mean-field potential in the latter case 
 is similar to the analytical prediction. 
On the other hand, the smaller gradient corresponding to ImQMD with  $\Delta x$ = 1.4 fm explains the lower oscillation frequency observed in this case. 
However, 
it is interesting to notice that, in spite of the fact that 
a gradient close to the analytical value is recovered in ImQMD for 
$\Delta x$ = 0.9 fm, 
oscillations are quite quenched in this case. This trend can be attributed
to the dominance of the damping effects associated with the
large fluctuation amplitude steming from the smaller Gaussian width.
In SMF calculations with 1 TP and in ImQMD-L, this effect is counterbalanced by the larger
value of the potential gradient (see Fig.\ref{gradz_SMF-ImQMD}(c) and (d)).


\begin{figure*}
\vskip 2.1cm
\includegraphics[scale=0.75]{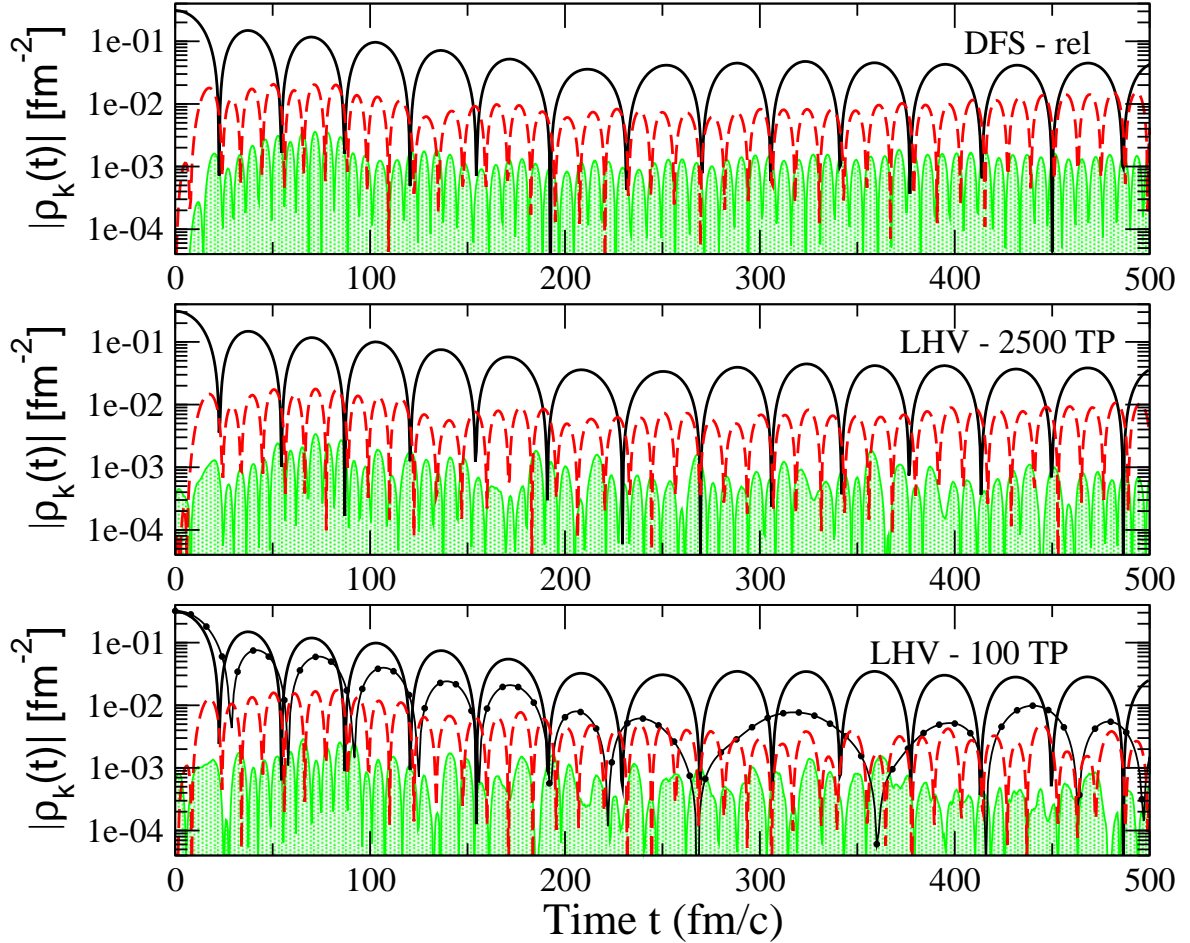}
\caption{Absolute strength $|\rho_k(t)|$ of different modes $n$ 
as a function of time: $n$ = 1 ((black) full line), $n$ = 3 ((red) dashed line), 
$n$ = 5 ((green) line with shading below). 
Results are shown for DFS calculations with relativitic kinematics (top panel)
and LHV calculations with 2500 TP (middle panel) and 100 TP (bottom panel)
per nucleon, initialized with mode $n=1$. 
The bottom panel also shows results from standard ImQMD calculations 
for $n$ = 1 ((black) line with dots).}
\label{modes_SMF}
\end{figure*}


\begin{figure*}
\vskip2.5cm
\includegraphics[scale=0.7]{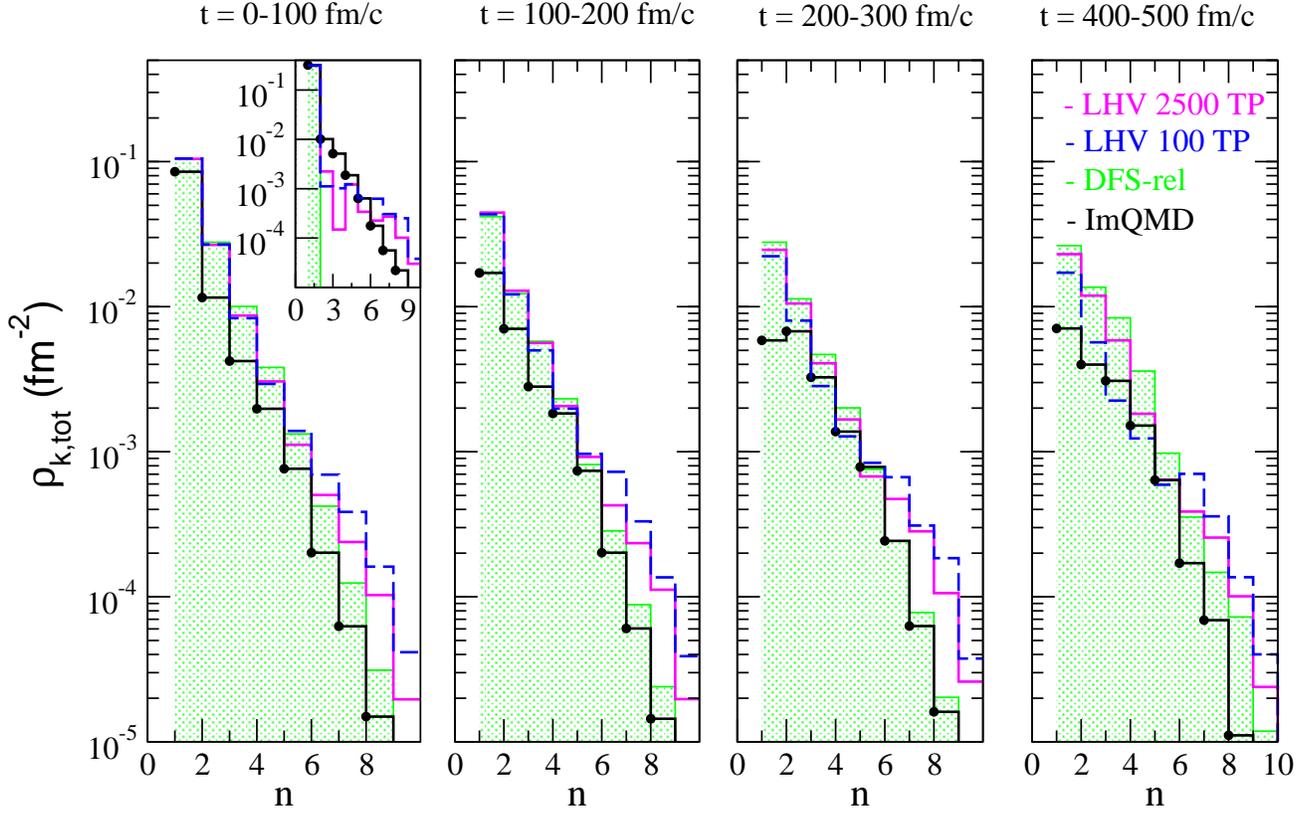}
\caption{The value of the strength $\rho_{k,tot}$ (see text), 
 for calculations initialized with mode $n=1$, as a function of the mode
number $n$. The different panels correspond to the average of $\rho_{k,tot}$ 
over the time interval indicated on the top.  Calculations are shown for
DFS calculations with relativistic kinematics (green shaded histogram), 
LHV calculations with 2500 TP ((magenta) full line) and 100 TP ((blue) dashed line)
per nucleon, and standard ImQMD calculations (full line with dots).
The inset in the first panel shows the distribution corresponding to the
initial time t = 0.
}
\label{modes_SMF2}
\end{figure*}
\subsection{Mode coupling}
\label{sec:selected_modes}
To understand the behavior observed for 
the strength of the initialized mode ($n=1$) already seen in Fig.\ref{oscill_all}, 
one has to consider the important coupling effects with other modes 
(with $n>1$), inducing anharmonicities.  These effects are
connected to the non-linear character of the Vlasov equation. 
This is shown in Fig.\ref{modes_SMF}, which displays the absolute value of the
strength of different modes as a function of time. 
DFS calculations in the relativistic
formulation are represented in the top panel. In particular, the figure shows the oscillations of the modes with 
$n=3$ and $n=5$, which are not present in the initial conditions but arise 
over time from the coupling to the $n = 1$ mode. The amplitude of these oscillations and their
time evolution is quite sensitive to the details of the mean-field potential
and its gradient.  
We also observe that the coupling to the other modes induces damping effects
in the $n=1$ mode, as also evident in Fig.\ref{DFS_strength_comp}. 
The DFS results are compared to LHV calculations with 2500 and 100 TP per nucleon in the middle and bottom panels of Fig.\ref{modes_SMF}, respectively. 
A nice agreement is observed for the calculations with 2500 TP, especially
for the dynamics of $n=1$ and $n=3$ modes. 
Employing 100 TP per nucleon, one can see that the dynamics of the $n=1$ 
mode is reasonably well preserved, whereas $n=3$ and
especially $n=5$ oscillations start to be dominated by a chaotic behavior 
attributable to numerical fluctuations. 
The bottom panel of Fig.\ref{modes_SMF} also shows standard ImQMD calculations
for the mode $n=1$. One can clearly appreciate the stronger damping and the 
loss of harmonicity at late times, as already discussed above. The modes
with $n=3$ and $n=5$ (not shown) are rather chaotic in this case. 

A deeper insight into mode coupling effects is obtained from Fig.\ref{modes_SMF2}, 
which shows the quantity $\rho_{k,tot}$ (see Sect. \ref{sec:hw_Fourier}), averaged over
the time interval indicated on the top of each panel, as a function of the 
node number $n$.  We recall that the system is initialized with $n=1$. 
The decreasing trend with mode number $n$ exhibited by DFS calculations is well reproduced by LHV
calculations with 2500 TP. When employing 100 TP, damping effects are visible 
at late times for the lower $n$ numbers. The overestimation, with respect
to DFS results, observed for the mode numbers $n \geq 4-5$ can be connected to
numerical fluctuations already present in the initial conditions (see the inset in the first panel).  
In ImQMD calculations, the amplitude of the modes $n \geq 2$ remains similar
to the initial value represented in the inset of the first panel, which is due 
to numerical density fluctuations.
The quenching of the modes with large $n$ can be
attributed to the density smearing effects associated with the Gaussian width. 
The mode $n=1$, which is excited in the initial conditions, is considerably damped, and approaches the amplitude associated with statistical density fluctuations (see Eq.(\ref{var})) already at t $\approx$ 200 fm/c.


\begin{figure*}
\vskip 2.8cm
\includegraphics[scale=0.6]{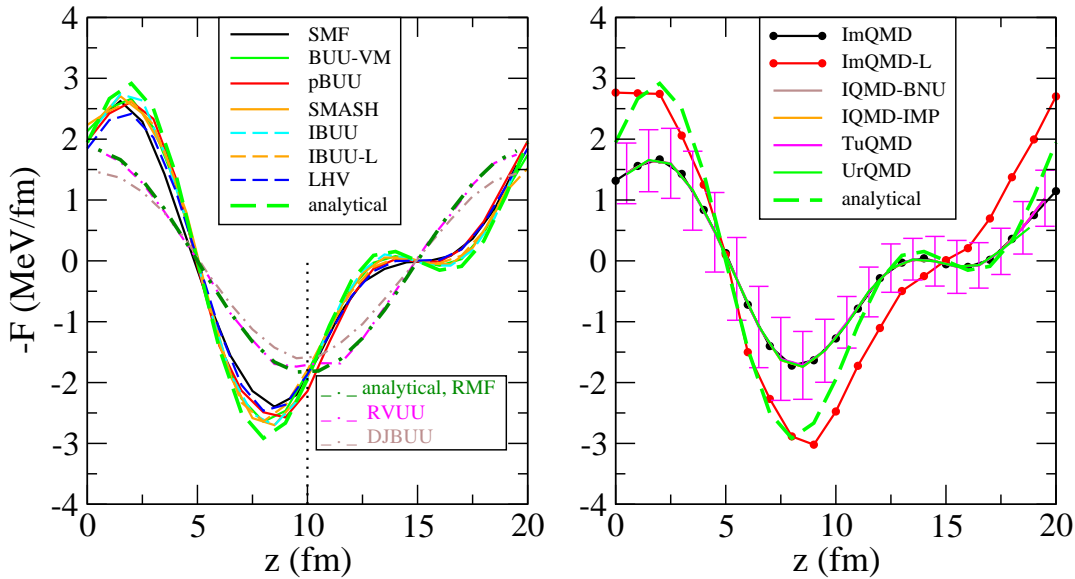}
\caption{
The gradient of the mean-field potential, at time t = 0, 
in the BUU-like (left panel) 
and QMD-like (right panel) codes, in comparison to the respective analytical 
result.  The QMD codes have used an identical initialization, the error bars represents 
the variance.
The dotted vertical line in the left panel indicated the central position 
of the box. 
}
\label{gradz_all} 
\end{figure*}


\section{Results of all participating codes}
\label{sec:all}


In this section we compare results of all the participating codes, using the numerical parameters recommended in the homework specification or chosen by the code owners. The focus is therefore on the more systematical 
similarities and differences
between the different types of codes and within each family.

According to arguments given above, we expect that the oscillation strongly depends on the behavior of  the potential  gradients as calculated in the codes. 
We therefore first show in Fig.\ref{gradz_all} the average gradients in $z$-direction for all the codes, at time t = 0.  The BUU codes give gradients close to the respective analytical results (note the different analytical prediction in the case of RVUU and DJBUU, 
as already explained in Sect. \ref{nucl_int}).  The QMD codes also give consistent results within this family, since they are using a common initialization, but generally lower than the analytical expectation. As discussed in Sec. \ref{sec:selected_gradZ}, this is due to the approximation used in evaluating the non-linear repulsive term of the force. This gives rise to generally lower frequencies of the oscillation for the QMD codes.
In the case of ImQMD-L, which is free 
from this approximation, the potential gradient is larger (in absolute value)
and becomes close to the analytical prediction.  

\begin{figure*}
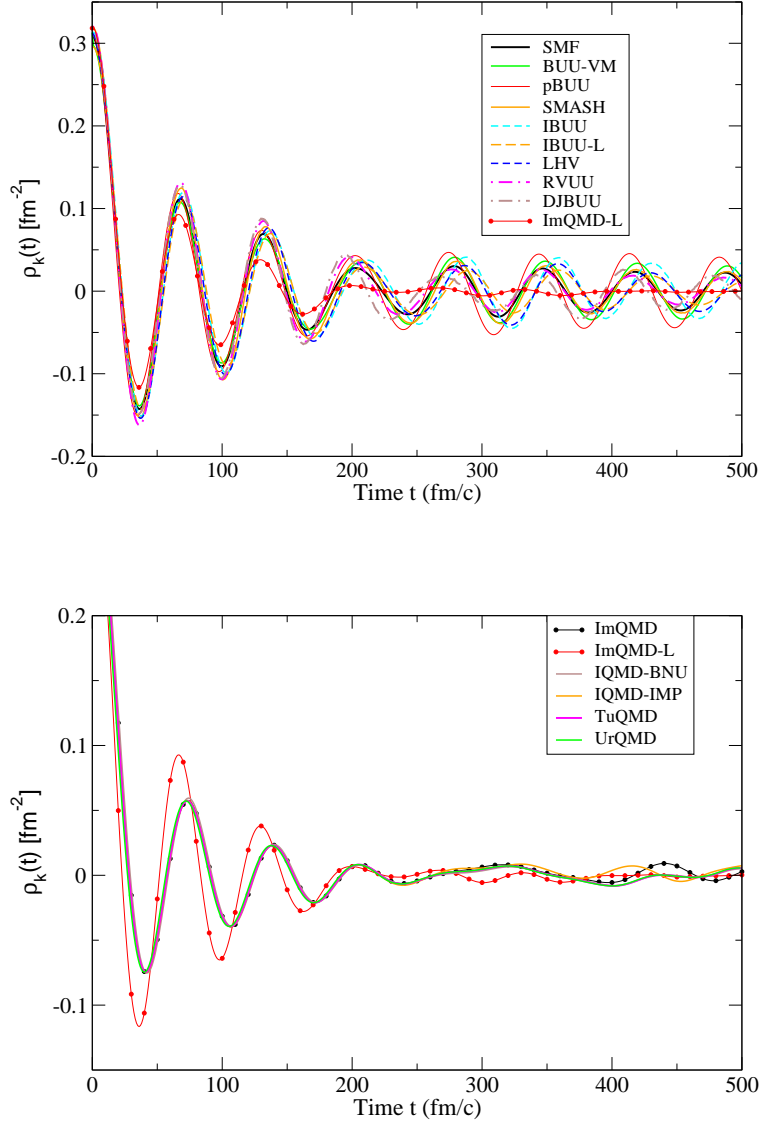

\includegraphics[scale=0.4]{fig14a.eps}
\vskip 1.3cm
\includegraphics[scale=0.4]{fig14b.eps}
\caption{
The strength function $\rho_k(t)$ for mode number $n$=1 is displayed as a function of time. Results are shown for BUU-like calculations (top panel, including
ImQMD-L calculations for comparison) 
and QMD-like calculations (bottom panel).
}
\label{FT_all}
\end{figure*}

\begin{figure*}
\vskip1.5 cm
\includegraphics[scale=0.60]{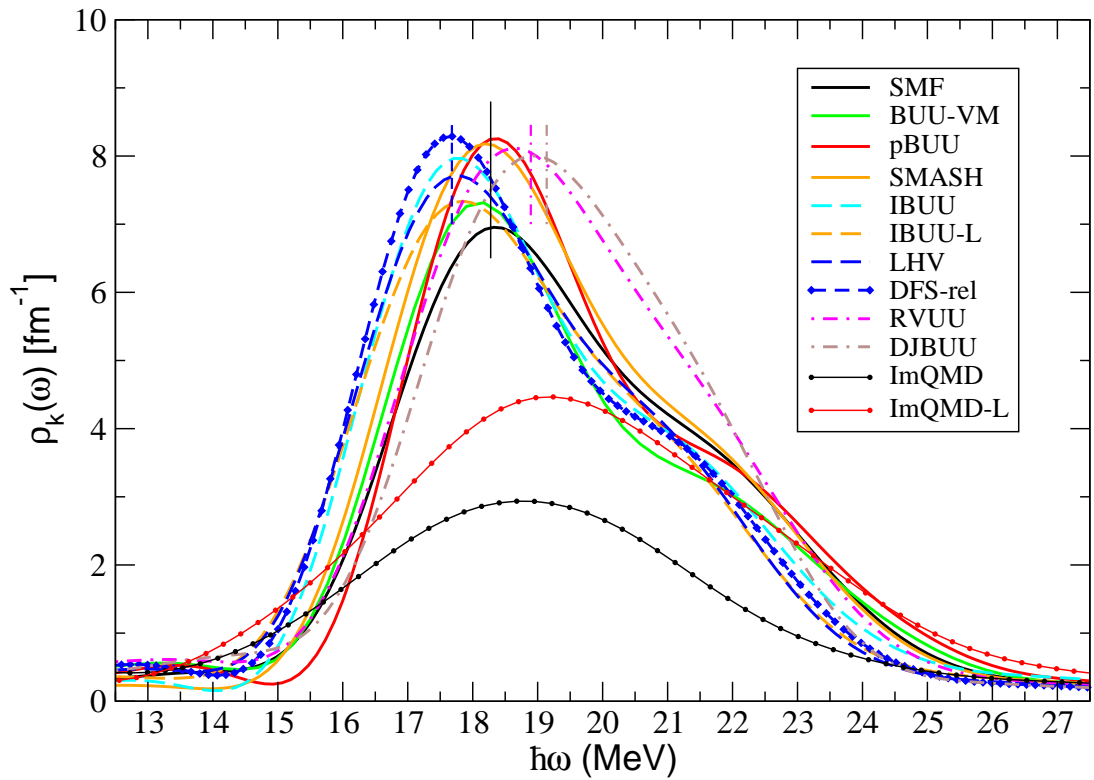}
\caption{Response function $\rho_k(\omega)$, i.e., Fourier transform with respect to space and time, of the averaged density distribution from 
BUU-like and two QMD-like calculations. Results are shown also for DFS calculations with relativistic kinematics.
The vertical lines indicate the analytical zero-sound energies for the different
code types, down-shifted by 2$\%$ (see text). 
} 
\label{response_BUU-QMD}
\end{figure*}

\subsection{Strength function}
The time evolution of the $n=1$ mode of the Fourier transform of the density oscillations, 
namely the strength function $\rho_k(t)$,  is displayed in Fig.\ref{FT_all} for all BUU-like (top panel) and
QMD-like (bottom panel) codes participating in the comparison. 
For the BUU-like codes, three main groups can be discerned (best visible
around t = 400 fm/c): slower oscillations
are seen for the codes of type ``rel'', namely IBUU, IBUU-L and LHV, compared
to the codes of type ``non-rel'' (SMF and BUU-VM), in line with the analytical expectations.  The covariant code SMASH exhibits similar oscillation 
frequencies, as compared to the codes of type ``non-rel'', whereas a slightly 
larger frequency is seen for pBUU, RVUU and DJBUU.
These features also reflect the analytical predictions of Table III, as it will be better illustrated in the next subsection. 
The amplitude of the oscillations at late times reflects the damping effects
associated with the number of test particles ($N_{TP}$ = 100) employed in the
calculations. The oscillations are less quenched for the codes which employed
a larger number of test particles in order
to preserve a good quality for the momentum distribution
(such as, for instance, BUU-VM, 
IBUU and pBUU).

As a general feature, the QMD-like codes in the lower panel show a stronger damping, which is consistent with the larger fluctuations in these codes, and also generally a smaller frequency, with respect to the analytical expectation (codes of type ``rel''), 
especially at early times, 
which is consistent with the reduced gradients in QMD, as seen in Fig.\ref{gradz_all}.
The frequency is higher for 
ImQMD-L, which is free from the approximation employed to evaluate the many-body term of the force in QMD. In this case, the early behavior of the Fourier 
transform coefficient is close to the results of the BUU-like codes, as one can 
appreciate from the top panel, where ImQMD-L results are also included.

\subsection{Response function}
\label{sec:response}
A compact presentation of the dynamical properties of the mean-field propagation is given by the response function, $\rho_k(\omega)$, which was introduced in Sect. \ref{sec:hw_Fourier} as the Fourier transform of the strength function with respect to time. This quantity is shown in Fig.\ref{response_BUU-QMD} for all the codes participating in the comparison. As the initial time, $t_{in}$, we consider the time 
instant of the first minimum of the Fourier transform coefficient $\rho_k(t)$ 
for each code.
This choice is motivated by the fact that, as explained above, the amplitude of the initial density perturbation impressed to the system is quickly quenched, by about a factor two, to reach the amplitude of 
the zero-sound collective mode. Thus zero-sound oscillations are more properly
investigated starting from the second peak in the time evolution (i.e., the
first minumum).    
To make the Fourier transform
with respect to time more meaningful, the function $\rho_k(t)$ is multiplied by the smearing function
$\cos^2[\pi(t-t_{in})/(2(t_{fin}-t_{in}))]$, so that at the final time the resulting product function goes to zero. More details about the sensitivity of the 
response function to $t_{in}$, and to smearing effects, are given in the 
Appendix \ref{resp_details}. 

 The response function should have a peak centered at the energy of the mode, and the width of the peak is a measure of the damping.  
Here the three groups of BUU-like codes already evidenced in Fig.\ref{FT_all} 
are nicely visible: for the codes of type ``non-rel'', i.e.,
BUU-VM and SMF, the peak energy is close to the one of 
pBUU and SMASH (these four codes
are denoted by full lines); the codes of type ``rel'', namely LHV, IBUU-L and
IBUU, have smaller frequency (dashed lines); the covariant codes RVUU and DJBUU (dot-dashed lines) exhibit a larger peak energy. 
This trend is in agreement with the analytical predictions given in Table III,
though the peak energies extracted from Fig.\ref{response_BUU-QMD} are slightly
smaller than the zero-sound energies. 
For instance, for codes of type ``non-rel'' one would expect a peak at the energy
$E = \hbar\omega$ = 18.65 MeV, which is slightly larger than the results of 
SMF (18.32 MeV) and BUU-VM (18.17 MeV). 
In the figure, this is evidenced by the
four vertical segments, which indicate the analytical zero-sound solutions 
corresponding (from the left to the right) to codes of type ``rel'', 
codes of type ``non-rel'', RVUU and DJBUU. The lines have been shifted down by 2$\%$ (to fit the 
DFS peak energy).     
This effect is mainly due to mode coupling; indeed it is observed also in the
case of the exact DFS calculations. 
The larger difference seen for RVUU could originate from the more significant
deviation from the Fermi statistics, with respect to the other BUU-like codes, 
as shown in Fig.\ref{moment_all}.  

The width of the response function reflects mode coupling and damping effects already discussed in the previous section.  In some cases, a shoulder is observed at energies 
larger than the peak energy, which can be attributed to the presence of 
non-linearities. Indeed, the latter effects tend to increase the oscillation
frequency, because of the coupling to modes with larger wave numbers.
However, it should be noticed that the width is also affected by numerical
ingredients, such as the final time considered and the smearing function
introduced to evaluate the response function (see the Appendix \ref{resp_details}). 
 
Since all QMD codes lead to an almost identical behavior for the time evolution
of $\rho_k(t)$, we show here only the results obtained for ImQMD and ImQMD-L. 
We clearly observe the quite large damping effects associated with the QMD-like
codes.  The strength is larger in the case of ImQMD-L, owing to the stronger
driving force in this case (i.e., to the larger gradient of the mean-field 
potential), that also leads to a higher peak energy, as compared to ImQMD.   
The peak energy observed for ImQMD is close to the BUU codes of type ``non-rel'', 
indicating that the reduced mean-field gradient values (see Fig.\ref{gradz_all}) mainly affect the early evolution of the system, that is excluded in our evaluation of the response function. 
   

\section{Discussions and Conclusion}
\label{sec:Discussion}

This paper continues evaluations of the robustness of transport-model predictions for heavy-ion collisions. One direction of these studies have been calculations in a periodic box, where ingredients of transport codes can be studied in separation and against results that are exact or that can be calculated more accurately with other methods.  After box investigations of elastic collisions with Pauli blocking \cite{comp2} and inelastic collisions without Pauli blocking \cite{comp3}, yielding Delta resonances and pions, we study here mean-field dynamics in a box, without collisions.  The system is initialized in terms of a standing density wave and the system evolution is followed with different participating codes 
using energy functionals that are made identical or similar.  Major transport codes from the two basic families, BUU and QMD, are included in this study, which
also partly account for relativistic effects in different approximations. We compare outcomes between the codes and to exact results in the small-amplitude limit and to numerical results for the evolution obtained in a more direct and accurate manner.  The comparisons include those of the strength function characterizing mode evolution and response function revealing how frequencies are tied 
to the modes.

\begin{figure*}
\vskip 1.9cm
\includegraphics[scale=0.7]{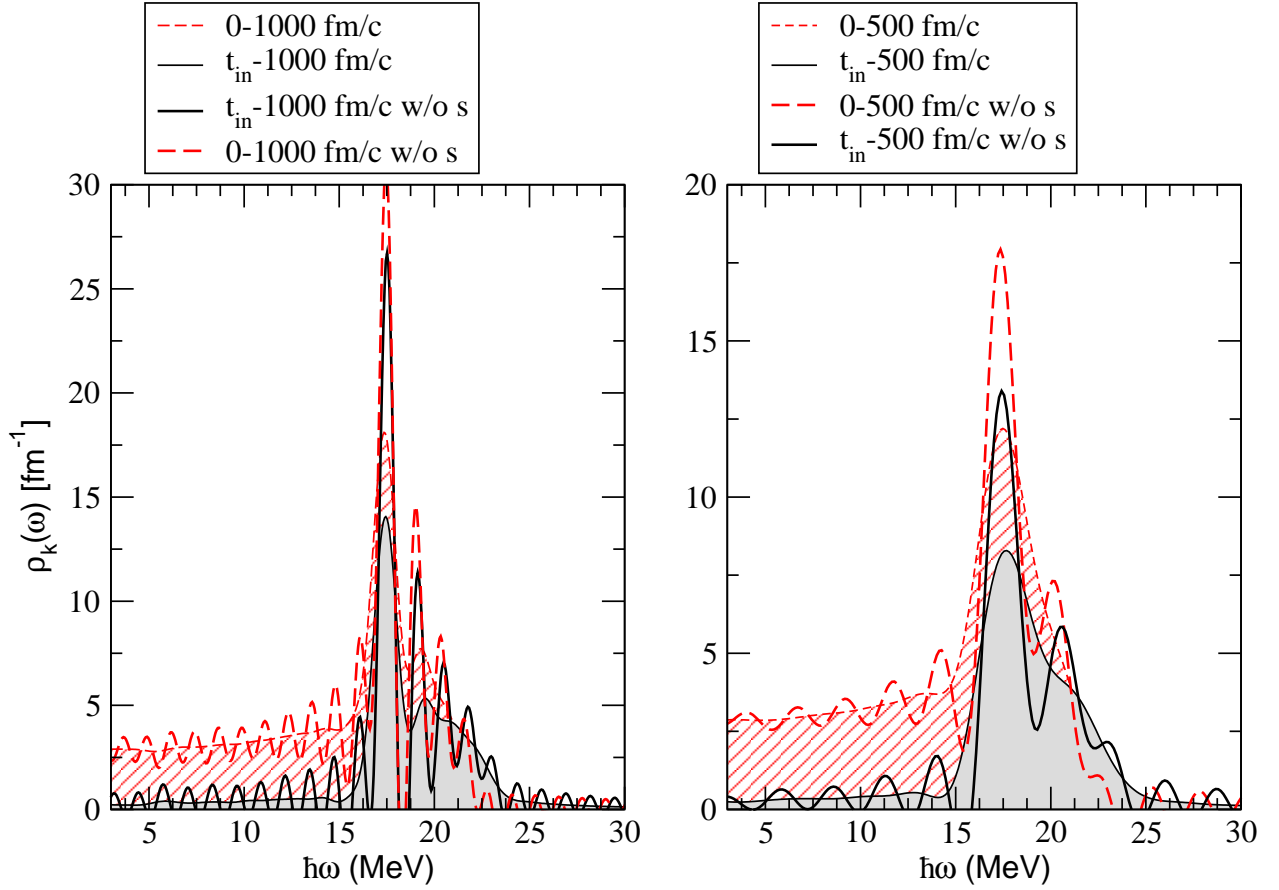}
\caption{
Response function $\rho_k(\omega)$, i.e. Fourier transform with respect to space and time of the averaged density
distribution, from DFS calculations with relativistic kinematics. Results are shown for different choices of the
time interval used for the Fourier transform with respect to time, and of the use of a smearing function or not. Results with the standard final time of  $t_{fin}$ = 500 fm/c are shown in the right panel, and for  $t_{fin}$ = 1000 fm/c in the left panel. Curves with the standard choice of the initial time (first minimum of the oscillation) are given by solid (black) curves, and  those with $t_{in}$= 0 by the dashed (red) curves. Results using the smoothing function are given by the thin curves (with shading below), and those without smearing (w/o s)  by the thick curves.
}
\label{DFS_time}
\end{figure*}
We find that we can generally understand consistencies and differences between the results of the codes.  
The differences among the codes and relative to near-exact results that persist include:  (1) relativistic effects that yield observable effects in the frequency of collisionless  mode; 
(2) approximations to the calculation of the non-linear terms of the force used in QMD codes that lead to noticeable differences in the 
frequency of the density oscillations even at early times, which can, however, be avoided in a lattice evaluation scheme; and
(3) the importance of damping effects generated by statistical or numerical 
fluctuations.
Indeed, the most noticeable differences in the results of the codes arise from the fluctuations inherent in the coarse phase space representation, which are characteristically different in BUU and QMD codes. They lead to a considerable damping of the modes, and in extreme cases also to frequency changes. We could show that by extremely extending the test particle number in BUU codes, we can come close to the near-exact results, as is to be expected. But already with more moderate numbers of test particle, as commonly used in heavy-ion calculations, the results compare well against the near-exact results. In QMD codes the damping is much stronger than in BUU, affecting also slightly the frequency. These findings do not make a statement about the validity of the two approaches, since the physical modeling is different: QMD codes attempt to put a reasonable amount of fluctuation already into the ansatz for the representation, while in BUU these would have to be included by an extra fluctuation term in the Langevin framework.

The findings for the long-term behavior are relevant to the uses of semiclassical transport in the studies of oscillations of isolated finite nuclei, including comparisons to quantum-mechanical calculations in TDHF and RPA \cite{Zheng2016,Burrello2019}. 
Most of these studies have been carried out on a case by case basis, rather than systematically.

In the context of heavy-ion collisions, it should be noted that we employ here unrealistically stiff equations-of-state (K=500 MeV), overemphasizing the strength of mean-field back-reacting forces and yielding more robust oscillations.  For more realistic incompressibilities, the oscillations would have been slower and far more strongly damped.
Here we have investigated the oscillations for the rather long time span of 500 fm/$c$, which contains many cycles. For a realistic heavy-ion collision, probably 
the time interval where the maximal density is reached,
of the order of a quarter or half of the cycle in Fig.\ref{FT_all} is relevant at intermediate energies.  Over such times, the results of the different codes are not so much different, as seen in Fig.\ref{FT_all}, and thus not too large effects from differences in mean-field integration are expected. 
Perhaps the stronger damping, and, in most cases, also the weaker forces of the QMD codes could lead to a weaker response and to systematically reduced flow effects, see for instance the comparative heavy-ion study of 
Ref.\cite{Xu2016}.  
However, in realistic studies of heavy-ion collisions momentum-dependent forces have to be used, unlike the forces used here, which could lead to larger differences in the mean-field propagation \cite{Dan-PLB,Zhang-PRC}.
The impact of momentum-dependent forces is presently investigated in comparative studies of box calculations and heavy-ion collisions.


\section*{Acknowledgements}

We warmly thank Janus Weil from FIAS, University of Frankfurt, 
who contributed with calculations performed with the code GiBUU in the early stages of this project. We also thank J. Maruhn and F. Matera for enlightening 
discussions. 

{M. Colonna acknowledges the supports from the European Unions Horizon 2020 research and innovation programme under Grant Agreement No. 654002.}
{P.~Danielewicz acknowledges support from the US Department of Energy under Grant No. DE-SC0019209.}
{H.~Wolter acknowledges support the Deutsche Forschungsgemeinschaft (DFG, German Research Foundation) under Germaany's Excellence Strategy - EXC-2094 - 390783311, ORIGINS;}
{J.~Xu acknowledges 
the support by the National Natural Science Foundation of China under 
Grant No. 11922514.}
{M. Kim and C.-H. Lee were supported by National Research Foundation of Korea (NRF) grants funded  by the Korean government (Ministry of Science and ICT and Ministry of Education) (No. 2016R1A5A1013277 and No. 2018R1D1A1B07048599).
Y. Kim was supported by the Rare Isotope Science Project of Institute for
Basic Science funded by Ministry of Science, ICT and Future Planning, and NRF of Korea (2013M7A1A1075764).
S. Jeon is supported in part by the Natural Sciences and Engineering Research Council of Canada.
Z. Zhang acknowledges the support by the National Natural Science Foundation of China under Grant No. 11905302.}
{Y. X. Zhang acknowledges the supports in part by National Science Foundation of China Nos. 11875323, 11961141003, 11475262, 11365004, National Key Basic Research Development Program of China under Grant No. 2018YFA0404404, and the Continuous Basic Scientific Research Project (No. WDJC-2019-13, BJ20002501).
}
{Y. J. Wang and Q.F. Li acknowledge the supports in part by National Science Foundation of China Nos. U2032145, 11875125, and 12047568, and the National Key Research and Development Program of China under Grant No. 2020YFE0202002.}
A. Sorensen and D. Oliinychenko received support through the U.S. Department of Energy, Office of Science, Office of Nuclear Physics, under contract number DE-AC02- 05CH11231 and received support within the framework of the Beam Energy Scan Theory (BEST) Topical Collaboration. A.~Ono acknowledges support from Japan Society for the Promotion of Science KAKENHI Grant Nos.\ 24105008 and 17K05432.
{ 
M.~B.~Tsang acknowledges the support by the US National Science Foundation
Grant No. PHY-1565546 and the U.S. Department of Energy under Grant Nos.  DE-SC0021235.
C. M. Ko acknowledges the support by the US Department of Energy under 
Award No. DE-SC0015266 and the Welch Foundation under Grant No. A-1358. 
B.A. Li and M.B. Tsang acknowledge the U.S. Department of Energy under Award Number DE-SC0013702. B.A. Li also acknowledges the CUSTIPEN (China-U.S. Theory Institute for Physics with Exotic Nuclei) under the US Department of Energy Grant No. DE-SC0009971.
L. W. Chen acknowledges the support by the National Natural Science Foundation of China under Grant No. 11625521 and National SKA Program of China No. 2020SKA0120300.
F. S. Zhang acknowledges National Natural Science Foundation of China under Grant No.11635003, 11025524, 11161130520.}



\appendix
\section{Details of Calculation of Response Function
}
\label{resp_details}
Here we want to illustrate the sensitivity of calculation of the response function to some technical choices mentioned in Sect.\ref{sec:response}: the time integration interval and the use of a ``smearing function''.
In Fig. 16 we show the results for the response function for the DFS 
calculation with relativistic kinematics, in the right panel for  $t_{fin}$=500 fm/c (the standard choice) and in the left panel for  $t_{fin}$=1000 fm/c.
In each panel we plot the results for our standard choice for  $t_{in}$ (first minimum
in the time evolution of  $\rho_k(t)$ ((black) solid curves)   and  for  $t_{in}$=0 ((red) dashed curves), and with smearing (standard choice, thin lines with shading below) or without smearing (thick lines, notation ``w/o s'' in the legend). Thus the thin (black) line 
corresponds to our standard choice, and the one in the right panel is the same curve as the one shown in Fig. 15 for DFS-rel.

One sees that all curves have a main peak, which is the peak of interest in Fig. 15 and can be compared 
to the frequency of the zero-sound oscillation. The position of the main peak is essentially unaffected by the various choices of the time interval and the smearing. Thus the conclusions of our response function analysis in Sect.\ref{sec:response} are robust against the choices for these technical parameters. However, it may still be of interest  to investigate the consequences of these choices on the shape of the response function, which is done in this appendix.

Without the smearing function the time-dependence of the strength function is cut off abruptly at the final time. Then we expect the appearance of structures at frequencies $\omega_n=n \pi / (t_{fin}-t_{in})$, with $n$ an integer. Indeed, we see these structures in the thick curves, which have half the spacing for the doubled time interval. Including the smearing function, the strength function smoothly goes to zero at the final time. Correspondingly these structures are smoothed out in the thin curves.

As discussed in Sec. \ref{structure}, with our choice of the initial momenta of the (test) particles, namely randomly in the local spherical Fermi surface, we do not initialize the proper momentum distribution of the physical zero-sound mode. 
Then the initial evolution of the system is characterized by a fast quenching
(by about a factor 2) of the initial density perturbation, which feeds
low-frequency components. 
Indeed we see in both panels, that low-frequency modes are excited when taking $t_{in}$=0,  and this is the more so if the total time interval is shorter, 
i.e., the fewer 
modes there are.

Finally one can observe that the asymmetry of the response function is due to the admixture of higher modes, which is mainly due to the non-linearity of the mean-field potential. These higher mode admixtures are clearly resolved without the smearing, and are mostly smoothed out with the standard choice of $t_{fin}$=500 fm/c in the right panel.

\end{document}